\documentclass[12pt]{article}

\usepackage{moreverb}
\usepackage{array}
\usepackage{tikz}
\usepackage{natbib}
\usepackage{booktabs}
\usepackage{amsmath}
\usepackage{amsfonts}
\usepackage{algorithm}
\usepackage{amssymb}
\usepackage{changepage}
\usetikzlibrary{arrows.meta}
\usepackage{graphicx} 
\usetikzlibrary{shapes,positioning}

\usepackage{fullpage}
\usepackage{setspace}
\usepackage[hyphens]{url}
\usepackage{moreverb}
\pdfoutput=1 
\newcommand\BibTeX{{\rmfamily B\kern-.05em \textsc{i\kern-.025em b}\kern-.08em
T\kern-.1667em\lower.7ex\hbox{E}\kern-.125emX}}

\usepackage{subcaption}

\usepackage{enumitem}

\usepackage{algpseudocode,algorithm}

\usepackage{enumitem}
\usepackage{booktabs}
\usepackage{multirow}
\usepackage{graphicx}
\usepackage{wrapfig}
\usepackage{bbm}
\usepackage{caption}
\usepackage{subcaption}
\usepackage{tikz}

\usepackage{ifthen}
\usetikzlibrary{shapes, shapes.geometric}
\usepackage{xcolor}

\newcommand{\pkg}[1]{{\normalfont\fontseries{b}\selectfont #1}}

\newif\ifDisplayChanges 
\DisplayChangesfalse

\ifDisplayChanges
\newcommand{\delete}[1]{\textcolor{red}{\sout{#1}}}

\newcommand{\redCommentToBeRemoved}[1]{\textcolor{red}{#1}}
\else
\newcommand{\delete}[1]{}

\newcommand{\redCommentToBeRemoved}[1]{}
\fi

\renewcommand{\P}{\mathbb{P}}
\newcommand{\one}{\mathbbm{1}}
\newcommand{\E}{\mathbb{E}}

\usepackage{xparse}
\ExplSyntaxOn
\NewDocumentCommand{\createbunch}{ m O{} m }
 {
  \clist_map_inline:nn { #3 } { \cs_new_protected:cpn { #2 ##1 } { #1 { ##1 } } }
 }
\ExplSyntaxOff

\createbunch{\mathbb}{R}
\createbunch{\hat}[hat]{L,U}
\createbunch{\mathcal}[cal]{D}

\newcommand{\setLowerBound}{\hat{L}_\alpha}
\newcommand{\setLowerBoundTwoSided}{\setLowerBound^{2\mathrm{S}}}
\newcommand{\setUpperBound}{\hat{U}_\alpha}
\newcommand{\setUpperBoundTwoSided}{\setUpperBound^{2\mathrm{S}}}
\newcommand{\threshold}{\tau}
\newcommand{\superlevelSet}{S_\threshold}
\newcommand{\superlevelSetCustomArg}[1]{S_{#1}}
\newcommand{\regressionFunction}{\eta}

\makeatletter
\algrenewcommand\ALG@beginalgorithmic{\normalfont}
\makeatother

\makeatletter
\let\@fnsymbol\@arabic
\makeatother

\RequirePackage[colorlinks,citecolor={blue!60!black},urlcolor={blue!70!black},linkcolor={red!60!black},breaklinks,hypertexnames=false,filecolor={red!60!black}]{hyperref}

\begin{document}

\title{Data-driven controlled subgroup selection in\\clinical trials}

\newcommand{\myand}{\hspace{0.1cm},\hspace{0.2cm}}
\newcommand{\myfinaland}{\hspace{0.2cm} and }
\newcommand{\myandlinebreak}{\hspace{0.1cm},\\}

\date{\today}
\author{Manuel M. M{\"u}ller\thanks{Statistical Laboratory, University of Cambridge, Cambridge, United Kingdom}\myand
Bj{\"o}rn Bornkamp\thanks{Novartis Pharma AG, Basel, Switzerland}\myand 
Frank Bretz\footnotemark[2]\protect\phantom{\footnotesize 2}\textsuperscript{,}\thanks{Medical University of Vienna, Center for Medical Data Science, Institute of Medical Statistics, Vienna, Austria}\myand
Timothy I. Cannings\thanks{School of Mathematics and Maxwell Institute for Mathematical Sciences, University of Edinburgh, Edinburgh, United Kingdom}\myandlinebreak
Wei Liu\thanks{School of Mathematical Sciences, University of Southampton, Southampton, United Kingdom}\myand 
Henry W. J. Reeve\thanks{School of Artificial Intelligence, Nanjing University, Nanjing, China}\myand
Richard J. Samworth\footnotemark[1]\myand
Nikolaos Sfikas\footnotemark[2]\myandlinebreak
Fang Wan\thanks{School of Mathematical Sciences, Lancaster University, Lancaster, United Kingdom}\myfinaland
Konstantinos Sechidis\footnotemark[2]}

\maketitle

\begin{abstract}
    Subgroup selection in clinical trials is essential for identifying patient groups that react differently to a treatment, thereby enabling personalised medicine. In particular, subgroup selection can identify patient groups that respond particularly well to a treatment or that encounter adverse events more often. However, this is a post-selection inference problem, which may pose challenges for traditional techniques used for subgroup analysis, such as increased Type I error rates and potential biases from data-driven subgroup identification. In this paper, we present two methods for subgroup selection in regression problems: one based on generalised linear modelling and another on isotonic regression. We demonstrate how these methods can be used for data-driven subgroup identification in the analysis of clinical trials, focusing on two distinct tasks: identifying patient groups that are safe from manifesting adverse events and identifying patient groups with high treatment effect, while controlling for Type~I error in both cases. A thorough simulation study is conducted to evaluate the strengths and weaknesses of each method, providing detailed insight into the sensitivity of the Type~I error rate control to modelling assumptions.
\end{abstract}


\renewcommand\thefootnote{}

\renewcommand\thefootnote{\fnsymbol{footnote}}
\setcounter{footnote}{0}

\section{Introduction}
\label{sec:intro}
\subsection{Motivation}
Randomised control trials (RCT) are the gold standard for evaluating the safety and efficacy of new treatments. Subgroup discovery and analysis are essential components of the drug development process, as they help identify how different patient groups respond to treatments. This analysis is critical for understanding variations in treatment effects or safety outcomes based on baseline characteristics, and it can lead to personalized treatment plans and improved patient outcomes \citep{lipkovich2017tutorial}.

In terms of safety, we would like to uncover subgroups of patients that are less susceptible to adverse events, allowing for the development of safer therapies. One example question that we would like to answer is: \emph{``Which participants can be considered low risk and do not require inpatient monitoring after dosing?''} This exact question is used in the recent Food and Drug Administration (FDA) guidance document titled ``Considerations for the Use of Artificial Intelligence To Support Regulatory Decision-Making'' \citep{FDA2025} and it will serve as one of the two motivating examples of our work. In terms of efficacy, subgroup discovery is also critical since it can reveal populations with high treatment effect, paving the way for more personalized and effective treatments. For example, we would like to answer questions such as: \emph{``Which subgroups of participants demonstrate more robust improvement in their condition due to the treatment?''} 

According to \cite{alosh2017tutorial}, there are three categories of efficacy-based subgroup analysis in drug development:
\begin{description}
\item[Inferential analyses:]  designed to make formal statistical inference about treatment effects within specific subgroups.  In these analyses, the subgroups of interest are prespecified before the study begins, often based on previous evidence and relevant factors. The sample size in the subgroups of interest is hence appropriately planned for reliable conclusions and adequate multiplicity adjustment will be predefined. 
\item[Exploratory analyses:]  conducted to generate hypotheses and identify potential subgroups, defined by baseline characteristics, that may benefit from a treatment by uncovering variations in treatment response. Due to their exploratory nature, these analyses provide considerable flexibility, and the findings can be purely data-driven. On the other hand, these analyses are often limited by their lack of statistical rigor and the potential for false positive results.
\item[Supportive analyses:] performed to determine if a treatment benefit from an RCT can be applied to pre-defined subgroups, ensuring treatment effect homogeneity. Forest plots are often used to visualise these effects across subgroups, and pre-specified cut-off values are used to define these subgroups.
\end{description}

Much of the current scepticism surrounding the practice of subgroup analysis can be attributed to methodology for exploratory analysis being used incorrectly to try to make inferential claims, for instance by post-hoc specification of the subgroups of interest for which an inferential analysis is conducted \citep{lagakos2006challenge, wang2007Statistics,altman2015subgroup,zhang2015subgroup}. As \cite{senn1997wisdom} put it: ``\emph{The equivalent in horse-racing would be, rather than betting on a named horse, merely to write `first past the post' on the betting slip and collect the winnings, whatever the result}''. However, as we shall see in this paper, recently proposed methodology allows for flexible data-driven selection of subgroups while still giving the strong statistical guarantees necessary for inferential analyses.

\subsection{A new way of approaching the problem of subgroup analysis}
In this paper, we focus on such novel approaches for data-driven subgroup selection. The framework we consider improves upon traditional approaches in two ways; firstly, it addresses the limitations of the categories presented above while retaining their desirable characteristics, and, secondly, it leads to stronger inferential claims for individuals falling into a selected subgroup. To illustrate the first point, we contrast our approach with the three categories above, highlighting how our methods integrate the strengths of each while overcoming their respective shortcomings.
\begin{description}
\item[Unlike inferential analyses,] where the subgroups traditionally need to be prespecified, our methods discover data-driven subgroups, while retaining Type I error guarantees. The only prespecified elements required are a (small) set of potential influential variables and the target response of interest.
\item[Unlike exploratory analyses,] where methods discover promising subgroups with different treatment effects from the overall population, often without Type I error guarantees, our methods perform targeted controlled analysis to identify subpopulations with desirable levels of treatment effects while controlling the chances of false discoveries.
\item[Unlike supportive analyses,] where the treatment effects are estimated in prespecified subgroupings (i.e., arbitrary cut-off values for continuous biomarkers), our methods learn the cut-off values (i.e., subgroups) from the data.
\end{description}
Concerning the second key aspect in which our framework differs from traditional approaches, it is worth mentioning that while we present the contributions in terms of the efficacy-based subgroup analysis, i.e., identify subgroups with high treatment effect, the general problem of subgroup selection that we focus on is broader and can be stated in any regression setting. Indeed, the methods we consider are suitable for settings where given a sample of covariate–response pairs, one considers the subgroup selection problem of identifying a subset of the covariate domain, where the regression function exceeds a predetermined threshold. In other words, we define the subgroup of interest by its worst-case effect on the subgroup. 

\begin{figure}
    \centering
    \begin{subfigure}{0.45\textwidth}
        \centering
        \includegraphics[width=\textwidth]{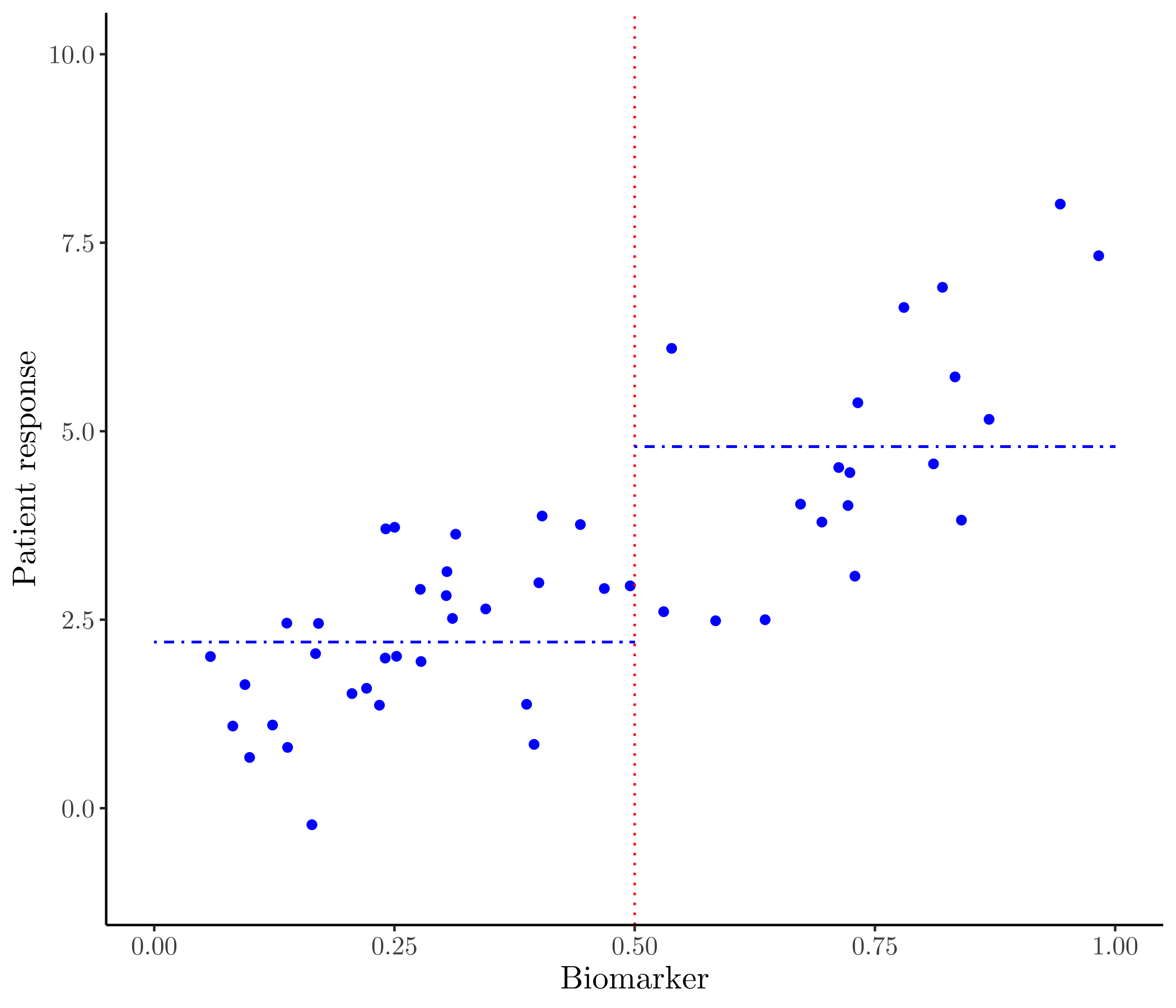}
        \caption{}
        \label{fig:into_subgroup_definition_conventional}
    \end{subfigure}
    \begin{subfigure}{.45\textwidth}
        \centering
        \includegraphics[width=\textwidth]{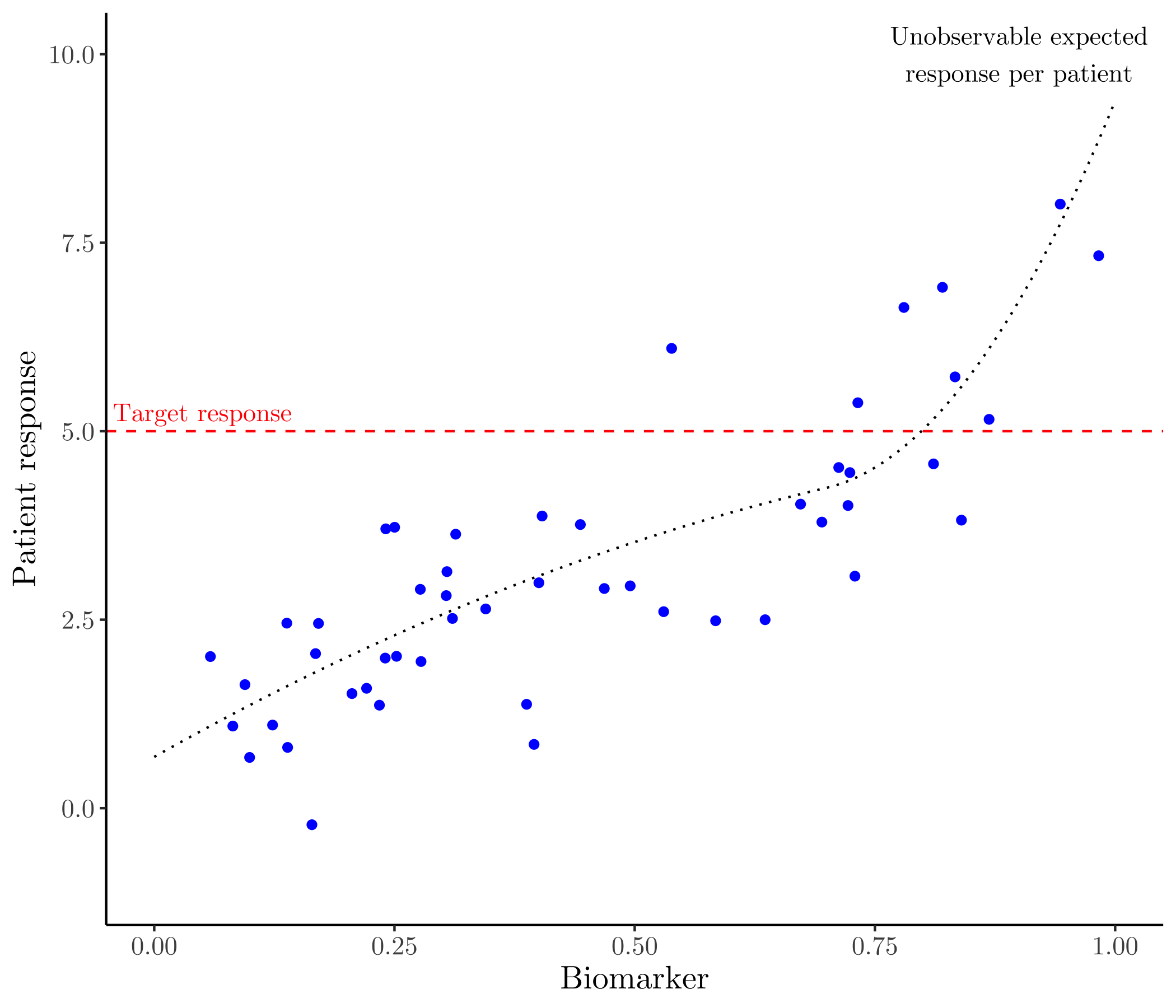}
        \caption{}
        \label{fig:into_subgroup_definition_pointwise}
    \end{subfigure}
    \caption{Toy example of different types of subgroup analysis. (a) One example of traditional subgroup analysis, where a pre-specified cut-off value is defined by a biomarker level (in this case $0.5$) and the average response is estimated in the two subgroups, (b) Our approach to the problem, where the user specifies a target response of interest (in this case $5.0$), and the method identifies the cut-off value from the data in such a way, that all members of the discovered subgroup have an expected response above $5.0$, with control of the Type I error.}
    \label{fig:intro_subgroup_definition}
\end{figure}

To illustrate how this idea differs from more conventional approaches, suppose we have a set of patients (blue dots in Figure \ref{fig:into_subgroup_definition_conventional}), and that for each patient we record a response (y-axis), and some baseline biomarker value (x-axis). For the left plot, assume that one has a pre-selected cut-off value for the biomarker (which is often the case in practice); in this example it is chosen as $0.5$. One could then categorize participants as biomarker-negative or biomarker-positive and derive the average group-specific response (represented by horizontal dash-dotted blue lines). This type of subgroup-specific effects is what is considered in many traditional approaches to subgroup analysis; in inferential analyses, their expectations are often the estimands of interest, in exploratory analyses, maximizing the difference between the two groups' average responses is a common goal, and in supportive analyses, these average effects are what is typically reported \citep{alosh2017tutorial}.

Our approach to the problem takes a different direction. Imagine we have the same patients, but this time we are interested in a specific target response; in Figure~\ref{fig:into_subgroup_definition_pointwise} this is illustrated with the target chosen as $5$. We aim to select patients who exhibit an expected response above this target while controlling for Type I error. To address this, we assume the existence of an (unobservable) population regression function, providing the expected response based on the biomarker value. In the figure, we can see that this regression function crosses the target response when the biomarker value exceeds 0.8, defining the subgroup we focus on in this paper. A key benefit of this notion of a subgroup is that it removes of any concern of further heterogeneity within the subgroup along the considered covariates, as every patient within the subgroup has an expected response exceeding the target threshold. This is in stark contrast to the conventional notion of a subgroup in Figure~\ref{fig:into_subgroup_definition_conventional}; through grouping into biomarker-negative and biomarker-positive subgroups and considering average effects, nothing can be said about an individual patient's expected response.

To formalize the statistical framework for this problem, consider a regression setting, with $X$ denoting the generic covariate vector and $Y$ the response. Writing $\eta(x) := \E(Y\mid X=x)$, $x\in \R^d$, for the population regression function, we are interested in identifying the superlevel-set $S_\tau := \{x\in\R^d:\eta(x)\geq\tau\}$ for a pre-specified threshold $\tau \in \R$. More specifically, we are considering the class of settings where the practitioner specifies a nominal limit $\alpha \in (0,1)$ on the Type I error rate and has access to a collection of independent and identically distributed covariate-response pairs $(X_1,Y_1),\ldots,(X_n,Y_n)$ following the same distribution as $(X,Y)$. Based on these observations, we seek to select a subset of $\R^d$ that is entirely contained within $S_\tau$ with probability at least $1-\alpha$. Equivalently, we may be interested in finding a subset of $\R^d$ that fully covers $S_\tau$ with probability at least $1-\alpha$, or we might seek to address these two tasks simultaneously. 
This framework may be extended from the pure regression setting (where a single outcome per patient is of interest) to the setting of treatment effects (i.e.~where the difference of outcomes across two interventions is of interest), where patients are characterized by triples $(X,T,Y)$, with $T \in \{0,1\}$ a binary control-vs-treatment indicator \citep{hernan2025causal}. Details are presented in Section~\ref{sec:methods_hte}.

\subsection{Related literature}
The framework we consider deviates from traditional methods in multiple ways. 

First and foremost, subgroup-specific effects are historically assessed by the \emph{average} effect on a subgroup, rather than by the \emph{worst case} effect on it, as we are doing here through the definition of $\superlevelSet$. This latter definition turns subgroup selection into the problem of conducting inference on the superlevel set of a regression function that has recently been addressed in the context of generalised linear models (GLMs) by \cite{wan2024confidence}, in smoothness-constrained nonparametric models by \cite{reeve2023optimal}, and in monotonicity-constrained nonparametric settings by \cite{mueller2024isotonic}. For the specific case in which the regression function gives a measure of the treatment effect, this framework was also considered in a linear model setting by \cite{ballarini2018subgroup} and in a univariate, monotonicity-constrained setting by \cite{stallard2024testing}. In a Bayesian context, this setting of targeting the superlevel set of a linear regression function giving the treatment effect has been considered by \cite{schnell2016bayesian}, for instance.

Beyond this strengthened guarantee, our framework can be distinguished from much of the existing literature through guaranteeing strong Type I error rate control of the selected set without overly restricting the candidate subsets to be investigated, reflecting recent trends towards principled methods of personalized medicine \citep{lipkovich2024modern}. In particular, in response to the dangers of post-hoc selection of subgroups, common guidance on subgroup analysis includes the pre-specification of a small number of biologically plausible subsets based on domain knowledge and correcting any reported results for multiplicity \citep{senn1997wisdom, rothwell2005subgroup, lagakos2006challenge, bretz2014multiplicity}. On the other side, the popularity of recently developed data mining tools for detecting subgroups of potential interest illustrate a desire to identify subgroups in a data-driven manner. Some noteworthy tools for such subgroup discovery include: \emph{STEPP} \citep{bonetti2000graphical, bonetti2004patterns, yip2016subpopulation}, a nonparametric smoothing technique to graphically assess treatment effect heterogeneity driven by a single feature which does not directly provide candidate subgroups but can be used to inform the choice thereof; and \emph{SIDES} \citep{lipkovich2011subgroup} (and its high-dimensional extension \emph{SIDEScreen} \citep{lipkovich2014strategies}), an iterative method for finding subgroups promising to exhibit large treatment effects with various ways of controlling their complexity. Moreover, a range of tree-based search techniques exist that seek to identify subgroups with differential treatment effects. 
For instance, \cite{su2008interaction, su2009subgroup} developed the technique of \emph{interaction trees} by building on the popular \emph{CART}-algorithm \citep{breiman1984cart} and using interaction effects between treatment indicator and the covariates as the splitting criterion. A shortcoming of these methods is their tendency to split along covariates with many distinct values, which is improved upon by \emph{GUIDE} \citep{loh2002regression,loh2015regression}. In a similar vein, \cite{seibold2016modelbased} propose model-based recursive partitioning, a very general method in which the splits are based on independence tests between the partitioning variables and the score function of a parametric model assumed for the effect of covariates and treatment on the primary endpoint. Another important tree-based method is given by \emph{causal trees} \citep{athey2016recursive} (as well as its random forest version \emph{causal forests} \citep{wager2018estimation}), which aims to give more accurate estimates of the subgroup-specific effect and use principled splitting strategies (for instance by being designed to reliably select subsets with differential treatment effects and on which the variance when estimating the treatment effect will be small in the case of causal trees).
The idea of these methods is to discover parts of the covariate domain that may be driving treatment effect heterogeneity. To confirm these subgroups while controlling the family-wise error rate, one would then need to use an independent fold of the data at a second stage. However, the concern with such splitting strategies is that they reduce power, as fewer observations are available at the confirmation stage, and failure to confirm candidates at this second step leads to no selection whatsoever. Recently, there has been growing interest in methods that replace such rigid two-stage procedures by multi-stage approaches allowing for flexible involvement of the user in the decision at each step \citep{duan2024interactive, cheng2025chiseling}.

Moreover, many recent developments for modelling the treatment effect have been made in the area of \emph{meta-learners} (see \cite{lipkovich2024modern} for an overview of recent developments), where the idea is to learn the treatment-specific outcome functions and use them to predict the unobservable treatment effect of each patient. This approach has been pioneered by \cite{foster2011subgroup} who dubbed this technique the ``Virtual Twins'' method. We investigate how these techniques can be combined with our methods through simulations and real-world applications.

\subsection{Our contribution}
Earlier we phrased two example questions for the problem of subgroup selection concerning safety and efficacy, and here is a refinement of these questions. Let us imagine we have a set of baseline covariates, and we are interested in answering the following questions: 
\begin{description}
    \item[Question 1 (safety):] \emph{``Can we select a subgroup of patients that have a guaranteed pre-specified probability of not manifesting an adverse event of interest, so that such patients do not require post-dose monitoring?''}
    \item[Question 2 (efficacy):] \emph{``Can we select a subgroup of patients whose treatment effect is above a given target level?''}
\end{description}
We aim to derive the subgroup while controlling for Type I error, in order to obtain confirmatory conclusions. These two questions will be our drivers to illustrate the utility of our novel methodology in drug development questions.

\citet[Prop.~1]{reeve2023optimal} proved that for these questions and when faced with continuous biomarkers, ensuring control of the Type~I error rate across all possible distributions will inevitably lead to trivial power (in the sense that an essentially empty set is selected with high probability) under a large class of data-generating mechanisms. As a result, some assumptions are necessary to yield non-trivial solutions to this problem. To this end, in this paper we present and extend two recently developed tools that build upon two different types of modelling assumptions: one based on generalised linear modelling (GLM) \citep{wan2024confidence} and another on nonparametric monotonicity constraints, which we refer to as Isotonic Subgroup Selection (ISS) \citep{mueller2024isotonic}. These methods are applicable in the regression setting, where one is given covariate-response pairs of the form $(X,Y)$, and can be extended to heterogeneous treatment effect settings, i.e.~where one has covariate-treatment-response triples $(X,T,Y)$, through inverse propensity weighting. We extend this approach further by building upon a doubly robust way to approximate individual treatment effects \citep{kennedy2023optimal}. Furthermore, we carry out an extensive simulation study, considering scenarios designed to closely resemble real clinical trial data as encountered in practice, including situations with both continuous and categorical covariates. The purpose of this analysis is two-fold: to assess the power of the methods, and to evaluate the extent to which they retain Type I error rate control under violations of their respective assumptions.

The paper is organized as follows. Section \ref{sec:application_clinical_trials} illustrates the application of our work in clinical trials, beginning with two case studies: one focusing on safety through subgroup selection for the absence of adverse reactions, and the other on efficacy through subgroup selection for desired treatment response. Section \ref{sec:methods} presents the methodological framework, detailing three approaches: subgroup selection using confidence sets for a level set in linear regression, in the context of multivariate isotonic regression, and for heterogeneous treatment effects. Section \ref{sec:simulation_study}  reports results from a comprehensive simulation study, where we explore the performance of the proposed methods and evaluate their operating characteristics under various scenarios. Section \ref{sec:application_clinical_trials_revisiting} revisits the clinical trial applications to demonstrate the practical implications of the methods, while Section \ref{sec:conclusions} concludes with a discussion of the broader impact of the methods presented in this paper and outlines possible future directions.

\section{Application in clinical trials}\label{sec:application_clinical_trials}

\subsection{Case study 1 (safety): Subgroup selection for the absence of adverse reactions} \label{sec:application_clinical_trials_AE}
The first case study focuses on subgroup selection based on adverse event risk in the context of a clinical program. Due to the program’s early termination and confidentiality constraints, only high-level methodological insights are shared. In early-phase clinical development, safety considerations are often complex due to the potential for a broad spectrum of adverse events, some of which may be serious in nature. These events can necessitate post-dose monitoring to ensure patient well-being, which introduces both clinical and operational challenges. 

A central question in this case study is whether specific patient subgroups can be identified as ``safe'', meaning those with a high likelihood of not experiencing the adverse event (AE) and therefore may not require post-dose monitoring. By modelling the occurrence of the AE as a binary outcome variable $ Y $, where $ Y = 1 $ indicates that the patient remained free of AEs and $ Y = 0 $ indicates that an AE occurred, we can formally pose the following research question.\\

\noindent \textbf{Research question for case study 1:}  
\emph{``Can we identify subpopulations for which the probability of not experiencing the adverse event, conditional on covariates, exceeds a predefined safety threshold $ \tau $, while maintaining control of the Type I error rate?''}

\subsection{Case Study 2 (efficacy): Subgroup selection for desired treatment response} \label{sec:application_clinical_trials_TEH}
The second case study focuses on a series of Phase III clinical trials evaluating a pharmaceutical compound targeting psoriatic arthritis (PsA), a chronic inflammatory condition affecting joints, entheses, and skin that can lead to impaired physical function and reduced quality of life \citep{future2}. Psoriatic arthritis severity is assessed using a range of clinical measures that capture different aspects of disease activity. Among these are composite endpoints, such as the binary ACR50 score, developed by the American College of Rheumatology \citep{felson1993american}. Complementing these are continuous endpoints like tender joint count, swollen joint count, patient global assessment, and physician global assessment, which provide a more granular evaluation of changes within specific disease domains. While our previous case study (Section \ref{sec:application_clinical_trials_AE}) focused on a binary outcome, in this one we aim to demonstrate the generality of the proposed methods by focusing on a continuous endpoint.

Cosentyx (secukinumab) is approved for treating adult patients with active psoriatic arthritis and has been evaluated in several clinical trials, i.e. FUTURE 1-5 \citep{future1,future2,future3, future4,future5}. Several studies have analyzed the FUTURE trials to investigate treatment effect heterogeneity with respect to the primary endpoint, ACR50 \citep{Sechidis2021, bornkamp2023predicting, Cardner2023, sechidis2025usingindividualizedtreatmenteffects}. With our work, we take a step further by focusing on the identification of patients who experience a treatment effect above a predefined threshold, and we formally pose the following research question.\\

\noindent \textbf{Research question for case study 2:} \emph{``Can we identify a subgroup of super responders, defined as a subpopulation in which the expected difference in the endpoint value at week 16 between treatment with Cosentyx and placebo exceeds a predefined threshold $\tau$, while maintaining control of the Type I error rate?''}

\section{Methodology}
\label{sec:methods}

We first recall the goal formulated in Section~\ref{sec:intro}. Given a target response $\tau \in \R$, we seek to recover the true subgroup $S_\tau = \{x\in\R^d: \eta(x)\geq\tau\}$, where $\eta$ is the population regression function so that $\eta(x)$ gives the expected response at $x\in \R^d$. Based on a set of independent and identically distributed covariate-response pairs $(X_1,Y_1),\ldots, (X_n,Y_n)$ sampled from a distribution with regression function $\eta$, we then seek to conduct inference on $S_\tau$. More specifically, in this section we give compact descriptions of methods that may be used to construct sets $\setLowerBound$, $\setUpperBound$, $\setLowerBoundTwoSided$ or $\setUpperBoundTwoSided$, such that
    \begin{align*}
        \P\bigl(\forall x \in \setLowerBound: x \in \superlevelSet\bigr) = \P\bigl(\setLowerBound \subseteq \superlevelSet\bigr) &\geq 1 - \alpha, \\
        \P\bigl(\forall x \in \superlevelSet: x \in \setUpperBound\bigr) = \P\bigl(\superlevelSet \subseteq \setUpperBound \bigr) &\geq 1 - \alpha, \\
        \P\bigl(\setLowerBoundTwoSided \subseteq \superlevelSet \subseteq \setUpperBoundTwoSided \bigr) &\geq 1 - \alpha,
    \end{align*}
for a nominal Type I error rate $\alpha \in (0,1)$, under suitable assumptions. Note that these properties mean that, with high probability,~$\setLowerBound$ (resp.~$\setLowerBoundTwoSided$) is a smaller set than $\superlevelSet$ and $\setUpperBound$ (resp.~$\setUpperBoundTwoSided$) is a larger set, so that they give lower and upper bounds on $\superlevelSet$ respectively, which is captured by our chosen notation. The superscript~$\mathrm{2S}$ refers to the property of having a two-sided bound. These confidence statements are what constitute our strong notion of Type~I error rate guarantee. 

The inferential guarantees considered here even allow for downstream applications under covariate shift. For instance, if $\hat{L}_\alpha \subseteq S_\tau$ (i.e.~we have not committed a Type~I error) and we have a patient with covariate values~$X_0$ such that $X_0 \in \hat{L}_\alpha$, then it follows that $X_0 \in S_\tau$ and hence $\eta(X_0)\geq \tau$, no matter whether $X_0$ has been drawn from the same distribution as $X_1,\ldots, X_n$ or has been picked in any other way. Corresponding conclusions can be made when instead $\hat{U}_\alpha$, $\setLowerBoundTwoSided$ or $\setUpperBoundTwoSided$ are considered. This is in stark contrast to the conclusions offered by conducting inference on the expected effect averaged over a subset (as compared in Figure~\ref{fig:intro_subgroup_definition}), where one can at best draw conclusions for the average value of $\eta(X_0)$ but also that only when $X_0$ is sampled from the same distribution as $X_1,\ldots, X_n$. In other words, in our framework, the covariate values in the training data $X_1,\ldots, X_n$ do not need to be representative for the target population to which the patient with values $X_0$ belongs. This is particularly important when considering that exclusion criteria imposed as part of the recruitment for clinical trials may cause the distribution of the trial population to deviate from that of the target population. Whenever such exclusion criteria only affect the covariate distribution, our statistical guarantees remain intact, a robustness typically not exhibited by conventional approaches.

\subsection{Subgroup selection in the context of multivariate linear regression}\label{sec:methods_glm}
One method due to \cite{wan2024confidence} for conducting controlled subgroup selection in generalised linear models (GLMs) and mixed effects models is based on techniques for constructing simultaneous confidence bands in linear regression. Here, we briefly describe their procedure for the setting of GLMs. 

Throughout, we suppose that the data generating mechanism may be accurately described by a GLM, i.e., that there exist a parameter vector $\beta \in \R^d$ and a strictly monotone link function $g:\R\rightarrow \R$, such that the covariate-response pair $(X,Y)$ satisfies
\[
g\bigl(\eta(x)\bigr) \equiv g\bigl(\mathbb{E}(Y\mid X=x)\bigr) = x^\top \beta
\]
for $x \in \R^d$. Without loss of generality, we assume $g$ is a strictly increasing function. A key observation is that the subgroup $\superlevelSet$ then has the property that 
\begin{equation}
    \superlevelSet = \bigl\{x\in\R^d:\eta(x) \geq \tau\bigr\} = \bigl\{x\in\R^d:g\bigl(\eta(x)\bigr) \geq g(\tau)\bigr\} = \bigl\{x\in\R^d:x^\top \beta \geq g(\tau)\bigr\}, \label{eq:superlevel_set_transform}
\end{equation}
so that subgroup selection may be achieved by conducting inference on $x^\top \beta$ simultaneously over all values of $x \in \R^d$. For that purpose, let $\hat{\beta}$ be an estimate of $\beta$ based on independent copies $(X_1,Y_1),\ldots, (X_n,Y_n)$ of $(X,Y)$, and assume that, asymptotically in $n$, $\sqrt{n} (\hat{\beta}-\beta)\sim \mathcal{N}_{d}(0,\Sigma)$. Further, let $\hat{\Sigma}$ be a consistent estimate of $\Sigma$ and let $\tilde{m}^2(x)$ be the estimated asymptotic variance of the linear part $\sqrt{n}\cdot x^\top\hat{\beta}$, that is, $\tilde{m}(x)=\sqrt{\widehat{\mathrm{Var}}\bigl(\sqrt{n}\cdot x^\top\hat{\beta}\bigr)}=\sqrt{x^\top\hat{\Sigma}x}$. For a given covariate region $K \subseteq \R^d$, \cite{wan2024confidence} discuss the computation of critical values $\hat{c}_1, \hat{c}_2 \in [0,\infty)$ such that 
\begin{align*} 
 \liminf_{n\rightarrow \infty}\P \left \{ \, x^\top \beta  \leq  x^\top \hat{\beta} + \hat{c}_1\tilde{m}( x ) \ \forall \, x \in K \, \right\} &\geq 1- \alpha \, ,  \\
 \liminf_{n\rightarrow \infty}\P \left \{ \, x^\top \beta  \geq  x^\top \hat{\beta} - \hat{c}_1 \tilde{m}( x ) \ \forall \, x \in K \, \right\} &\geq 1- \alpha \, ,   \\
 \liminf_{n\rightarrow \infty} \P \left \{ \, x^\top \hat{\beta} - \hat{c}_2 \tilde{m}( x ) \leq x^\top \beta  \leq  x^\top\hat{\beta} + \hat{c}_2 \tilde{m}( x ) \ \forall \, x \in K \, \right\} &\geq 1- \alpha \, .
\end{align*}
Combining this with the observation in~\eqref{eq:superlevel_set_transform}, we let 
\begin{align*} 
\hat {U}_\alpha & :=  \left \{ \, x \in K : \ x^\top \hat{\beta} + \hat{c}_1 \tilde{m}( x ) \ge g(\tau)  \, \right\} \, ,   \\
 \hat {L}_\alpha & :=  \left \{ \, x \in K : \ x^\top\hat{\beta} - \hat{c}_1 \tilde{m}( x ) \ge g(\tau)  \, \right\} \, ,   \\
\setUpperBound ^{2\mathrm {S}} & :=  \left \{ \, x \in K : \ x^\top\hat{\beta} + \hat{c}_2 \tilde{m}( x ) \ge g(\tau)  \, \right\} ,  \ 
\setLowerBound ^{2\mathrm {S}} := \left \{ \, x \in K : \ x^\top \hat{\beta} - \hat{c}_2 \tilde{m}( x ) \ge g(\tau)  \, \right\} .  
\end{align*}
Indeed, we then have
\begin{align*} 
 \liminf_{n\rightarrow \infty}\P \left \{ \, \superlevelSet \subseteq \setUpperBound \, \right\} &\geq 1- \alpha \, ,  \\
 \liminf_{n\rightarrow \infty}\P \left \{ \, \setLowerBound \subseteq \superlevelSet \, \right\} &\geq 1- \alpha \, ,   \\
 \liminf_{n\rightarrow \infty} \P \left \{ \, \setLowerBound ^{2\mathrm {S}} \subseteq \superlevelSet \subseteq \setUpperBound^{2\mathrm {S}} \, \right\} &\geq 1- \alpha \, .
\end{align*}
The amount by which the asymptotic coverage probability exceeds $1- \alpha$ depends on the data generating distribution, though the construction of $\hat{c}_1$ due to \cite{wan2024confidence} ensures that the first two statements hold with equality for certain worst-case distributions.

In the special case of a normal linear model, i.e., where $g(t) = t$, so that $\eta(x) = x^\top \beta$, and $Y = X^\top\beta + \varepsilon$ for $\varepsilon\sim\mathcal{N}(0,\sigma^2)$, independent of $X$, the stated Type I error control also holds for finite sample sizes when using ordinary least squares for estimating $\beta$ and $\sigma$. While the two-sided construction may be conservative in this setting, more recent work due to \cite{wan2025exact} provides an alternative construction of $\hat{c}_2$, such that $1-\alpha$ is achieved exactly for certain distributions when $x$ consists of the intercept and one covariate.

\subsection{Subgroup selection in the context of multivariate isotonic regression}\label{sec:methods_iss}
\cite{mueller2024isotonic} introduce an approach for subgroup selection in the context of multivariate isotonic regression based on martingale tests and multiple testing procedures for logically structured hypotheses. It is implemented in the \texttt{R}-package \textbf{ISS} \citep{Mueller2023ISS_Code}. This method has for instance been used to identify adults at risk of depression based on adverse childhood experiences \citep{zhang2025acescorereplicablecombinations}. Here we give a high-level summary of the method, beginning with the univariate case for simplicity of exposition.

As before, let $(X,Y)$ denote a generic covariate-response-pair such that $\eta(x) = \E(Y\mid X=x)$ is increasing for $x\in\R$. If instead $\eta$ were decreasing, then one could consider the transformed covariate-response-pair $(-X,Y)$ for which one would then have an increasing relationship. Hence, in everything that follows, one could replace the assumption of $\eta$ being increasing with the assumption that $\eta$ is monotone with known direction (either increasing or decreasing). Given access to independent copies $(X_1, Y_1), \ldots, (X_n,Y_n)$ of $(X,Y)$, first fix $x_0$ to be the largest value among $X_1, \ldots, X_n$. In order to construct a test of the null hypothesis\footnote{As shown by \cite{mueller2024isotonic}, the described procedure will be valid for any realisation of $X_1, \ldots, X_n$, so that we may consider them as fixed throughout.} that $\eta(x_0) < \tau$, one defines a test sequence, where each element $T_k$, $k = 1, \ldots, \#\{i:X_i \leq x_0\}$, comprises the $k$ responses whose covariates are closest to $x_0$. The null hypothesis is rejected if for any $k$ the test statistic $T_k$ exceeds a critical threshold $c_k$ depending on $\alpha$. The threshold sequence $c_k$ can be chosen such that this procedure yields a provably valid test of the null hypothesis, as illustrated in Sections~\ref{sec:methods_ISS_pvalue_binary} and~\ref{sec:methods_ISS_pvalue_quantile}. 
 The reason for choosing such a sequential procedure is that under the alternative hypothesis that $\eta(x_0) \geq \tau$, the set of covariates $X_1, \ldots, X_n$ having corresponding regression function values $\eta(X_1), \ldots, \eta(X_n)$ at least at level $\tau$ is unknown. Supposing the hypothesis $\eta(x_0) < \tau$ is rejected at level $\alpha$, the covariate-response pair with covariate value $x_0$ is removed from the data set, and the above testing procedure is repeated with $x_0$ set to the next-largest covariate value. If at any point the null hypothesis $\eta(x_0) < \tau$ is not rejected at level $\alpha$ (or if all data points have been removed from the data set), then $x^*$ is taken to be the last value of $x_0$ for which the null was rejected and the procedure outputs $\hat{L}_\alpha := [x^*, \infty)$. If no null hypothesis is rejected, then $\hat{L}_\alpha$ will be the empty set. An upper confidence bound $\hat{U}_\alpha$ on $S_\tau$ can be obtained in a similar way, by calculating~$\hat{L}_\alpha$ on transformed data $(-X_1,-Y_1), \ldots, (-X_n, -Y_n)$ with a transformed threshold $-\tau$, where the resulting set is denoted $\Tilde{L}$, and then letting $\hat{U}_\alpha := \mathbb{R} \setminus \{x \in \mathbb{R}: -x\in \Tilde{L}\}$. Two-sided confidence sets can be obtained by letting $\hat{U}^{\mathrm{2S}}_\alpha := \hat{U}_{\alpha/2}$ and $\hat{L}^{\mathrm{2S}}_\alpha := \hat{L}_{\alpha/2}$.

The procedure above consists of two nested loops: in the outer loop, the choice of $x_0$ is varied from the largest value in $X_1, \ldots, X_n$ to the smallest, and the null hypothesis $\eta(x_0) < \tau$ is formulated at each step, while in the inner loop, an associated test statistic is calculated by iterating through values of $k$. While the critical thresholds $c_k$ need to account for the multiplicity in the sequential structure of the test, no such multiplicity correction is present in the outer loop, where the test may be carried out at level $\alpha$ each time. The reason for this lies in the fact that the procedure terminates immediately if at any step the null hypothesis is not rejected, so that family-wise error rate control is guaranteed \citep{hsu1999stepwise, westfall2001optimally}.

With additional work, a similar procedure can be developed for the multivariate setting. More specifically, since here $X_1, \ldots, X_n$ are vectors in $\R^d$ for some $d \in \mathbb{N}$, there is no natural total ordering: we may have two covariate vectors, neither of which is larger than the other in all components at once. Such vectors are said to be \emph{incomparable points} or to be forming an 
\emph{antichain}; this will play a role in Section~\ref{sec:methods_iss_categorical}, where categorical variables are discussed. For the construction of the test sequence, this issue may be remedied by considering only responses whose covariate vectors are smaller in every component than the vector $x_0 \in \R^d$ and letting them enter the test sequence $T_k$ in the order of the covariate vectors' supremum-norm distance from~$x_0$.  Designing an appropriate multiple testing procedure to constitute the outer loop described above is a little more involved, but 
\cite{mueller2024isotonic} propose a method that builds on the idea of constructing a directed acyclic graph representing a logical structure between the null hypotheses to be tested.  The $\alpha$-budget is then iteratively re-assigned to these hypotheses to again guarantee family-wise error rate control; see also work by \cite{bretz2009graphical}, \cite{goeman2010sequential}, and \cite{meijer2015multiple}.

We now turn to specific constructions of $p$-values that can be applied as part of the isotonic subgroup selection algorithm for particular types of responses.

\subsubsection{Binary responses}\label{sec:methods_ISS_pvalue_binary}
If the responses are binary, i.e.~they take values $0$ or $1$, then $Y_i$ given $X_i=x$ follows a Bernoulli distribution with the probability of observing a $1$ given by $\eta(x)$. In line with the general ideas outlined above, given a point $x_0 = \bigl(x_0^{(1)}, \ldots, x_0^{(d)}\bigr) \in \R^d$ at which we wish to test the null hypothesis $\eta(x_0) < \tau$, we propose to order data points whose covariate value in each dimension is at most that of~$x_0$ according to the supremum-norm distance from $x_0$.  Denote these ordered covariate vectors by $X_{(1)},\ldots,X_{(n(x_0))}$, for $n(x_0) := \#\bigl\{i:X_i^{(j)}\leq x_0^{(j)} \text{ for all }j\in \{1,\ldots,d\}\bigr\}$, and let $Y_{(1)}, \ldots, Y_{(n(x_0))}$ be the correspondingly ordered responses.  Then the sequence $T_k := \sum_{j=1}^k Y_{(j)}$ for $k \in \{1,\ldots, n(x_0)\}$ tracks a cumulative count of the number of $1$s, starting with the points expected to be most indicative of the value of $\eta$ at $x_0$.  Under the null hypothesis, every $Y_{(j)}$ takes the value $1$ with probability less than $\tau$, so $T_k$ should not grow to quickly (in particular, on average no faster than $k\tau$).  This enables the construction of critical thresholds~$c_k$ such that the null hypothesis should be rejected whenever $T_k > c_k$ for some $k$. \cite{mueller2024isotonic} show that this idea may be formulated as a $p$-value, via the formula
\[
\hat{p}_{\mathrm{Bin}} := \min\Bigl(1, \min_{k\in\{1,\ldots, n(x_0)\}}\frac{\tau^{T_k}(1-\tau)^{k-T_k + 1}}{B(1-\tau; k - T_k+1, T_k+1)}\Bigr),
\]
where the function $B(z;a,b) := \int_0^zt^{a-1}(1-t)^{b-1} \, dt$ for $z \in [0,1]$ and $a,b > 0$, denotes the incomplete beta function.  As an example, suppose that $\alpha = 0.1$ and $\tau = 1/2$.  Then $\hat{p}_{\mathrm{Bin}} \leq 0.1$ can only occur once $k\geq 5$, and only then if only ones have been observed, i.e.~$T_5 \geq 5 = c_5$.  Alternatively, the null hypothesis is rejected if either $T_9 \geq 8 = c_9$ or $T_{13} \geq 11 = c_{13}$ or $T_{16} \geq 13 = c_{16}$ or $T_{19} \geq 15 = c_{19}$ and so on, where the proportion of $1$s necessary to have been observed by step $k$ in order to reject the null, decreases towards $\tau = 1/2$. For instance, at $k=100$, observing $T_{100} \geq 64$ suffices for a rejection of the null hypothesis that $\eta(x_0) < 0.5$.

\subsubsection{Continuous responses}\label{sec:methods_ISS_pvalue_quantile}
There are different ways in which continuous responses can be handled. For instance, \cite{mueller2024isotonic} propose a way of constructing $p$-values targeting the specific case in which the responses may be assumed to follow a homoscedastic Gaussian distribution around their conditional expectation. An approach relying even less on distributional assumptions is based on the idea of considering quantiles of the response distribution. Here, the idea is to take the original responses $Y_i$ and dichotomize them into $Y^*_i := \one\{Y_i > \tau\}$. Thus, when applying the $p$-value for binary responses presented in Section~\ref{sec:methods_ISS_pvalue_binary} to the data $(X_i, Y^*_i), i \in \{1,\ldots,n\}$, a point $x_0 \in \R^d$, and a new threshold $\tau^*$, we yield a $p$-value for the null hypothesis that $\E(Y^*_i\mid X_i=x_0) = \P(Y_i > \tau \mid X_i=x_0) \leq \tau^*$. In particular, if we choose $\tau^* = 1/2$, we are testing the null hypothesis that the conditional median of $Y_i$ (given $X_i = x_0$) is less than $\tau$. As a result, under the general assumption that the distribution of the responses is symmetric about their conditional expectation, this approach also leads to recovering $S_\tau$ with the same Type I error rate guarantees, in the setting of continuous responses.

As highlighted in \cite{mueller2024isotonic}, further methods exist, e.g.~in the setting of bounded or sub-Gaussian response distributions.

\subsubsection{Categorical covariates}\label{sec:methods_iss_categorical}

The method presented at the beginning of Section~\ref{sec:methods_iss} can be applied equally well when categorical covariates are present, but does require that the direction of monotonicity is known.  We now present an effective modification for settings where this direction may be unknown, or where the monotonicity in the other covariates may hold only separately within each category and not jointly in all variables simultaneously.

\begin{figure}
    \centering
    \resizebox{0.8\textwidth}{!}{\definecolor{myorange}{HTML}{E69F00}
\definecolor{myblue}{HTML}{56B4E9}
\definecolor{mygreen}{HTML}{009E73}
\definecolor{mypink}{HTML}{CC79A7}
\newcommand{\pointsizeleft}{10pt}
\newcommand{\pointsizeright}{8pt}

\begin{tikzpicture}

\draw[->] (-0.5,0) -- (11,0) node[right] {$x_1$};
\draw[->] (0,-0.5) -- (0,11) node[above] {$x_2$};

\foreach \x/\y/\group in {
2/3/a,
4/7/b,
6/2/c,
5/5/a,
3/8/b,
8/4/c,
4/2/b,
8/6/b,
3/6/c,
8/3/a,
} {
  \ifthenelse{\equal{\group}{a}}{
    \node[draw=myblue, fill=myblue, shape=circle, minimum size=\pointsizeleft, inner sep=0pt] at (\x,\y) {};
  }{
  \ifthenelse{\equal{\group}{b}}{
    \node[draw=mypink, fill=mypink, shape=rectangle, minimum size=\pointsizeleft, inner sep=0pt] at (\x,\y) {};
  }{
  \ifthenelse{\equal{\group}{c}}{
    \node[draw=myorange, fill=myorange, shape=diamond, minimum size=1.4*\pointsizeleft, inner sep=0pt] at (\x,\y) {};
  }{}}}
}

\draw[->, very thick] (11.5,5.5) -- (13,5.5) node[right] {};

\end{tikzpicture}
\begin{tikzpicture}

\draw[->] (-0.5,0) -- (11,0) node[right] {$\xi_1$};
\draw[->] (0,-0.5) -- (0,11) node[above] {$\xi_2$};

\foreach \x/\y/\group in {
2/3/a,
4/7/b,
6/2/c,
5/5/a,
3/8/b,
8/4/c,
4/2/b,
8/6/b,
3/6/c,
8/3/a,
} {
  \ifthenelse{\equal{\group}{a}}{
    \node[draw=myblue, fill=myblue, shape=circle, minimum size=\pointsizeright, inner sep=0pt] at (\x/3 + 1,\y/3 + 8) {};
  }{
  \ifthenelse{\equal{\group}{b}}{
    \node[draw=mypink, fill=mypink, shape=rectangle, minimum size=\pointsizeright, inner sep=0pt] at (\x/3 + 4,\y/3 + 4) {};
  }{
  \ifthenelse{\equal{\group}{c}}{
    \node[draw=myorange, fill=myorange, shape=diamond, minimum size=1.4*\pointsizeright, inner sep=0pt] at (\x/3 + 7,\y/3 + 1) {};
  }{}}}
}

\draw[->] (0.8,8) -- (4,8) node[right] {$x_1$};
\draw[->] (1,7.8) -- (1,11) node[above] {$x_2$};

\draw[->] (3.8,4) -- (7,4) node[right] {$x_1$};
\draw[->] (4,3.8) -- (4,7) node[above] {$x_2$};

\draw[->] (7.0,0.8) -- (10.2,0.8) node[right] {$x_1$};
\draw[->] (7.2,0.6) -- (7.2,3.3) node[above] {$x_2$};

\end{tikzpicture}}
    \caption{The antichaining approach discussed in Section~\ref{sec:methods_iss_categorical}. Each dot represents a covariate vector, where differing symbols (purple squares versus orange diamonds versus blue circles) indicate different group memberships, and where groups may be defined by either one or multiple categorical variables.}
    \label{fig:antichaining}
\end{figure}
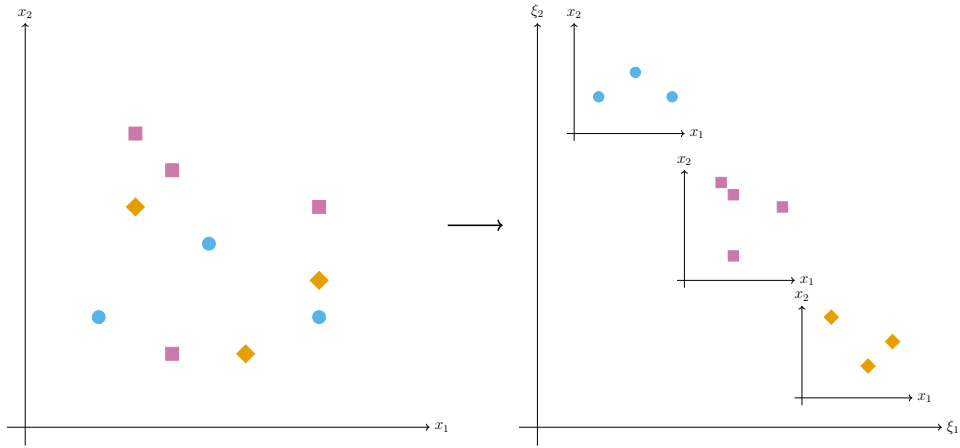

Suppose that there are $L$ levels of (combinations of) categorical variables about which we do not wish to pre-specify a monotonicity direction.  We introduce a variable $\xi_1$ taking values in $\{1,\ldots, L\}$ to index these levels, and also include the auxiliary variable $\xi_2 := -\xi_1$. Figure~\ref{fig:antichaining} illustrates this idea. In the left plot, $10$ covariate points are illustrated, each of which belongs to one of three groups and contains information on the covariates $x_1$ and $x_2$. The right plot shows their placement in a four-dimensional space corresponding to $x_1$, $x_2$, $\xi_1$, $\xi_2$, where the construction of the last two variables  leads to points of different groups forming a so-called \emph{antichain}. This \emph{antichaining approach} results in a setting in which neither point of any pair of two points belonging to distinct groups is larger than the other in every dimension. In addition, it has the benefit that no notion of distance needs to be specified between the levels of the categorical variables.

\subsection{Subgroup selection for heterogeneous treatment effects}\label{sec:methods_hte}
Meta-learners are advanced machine learning models that can be used to estimate the Conditional Average Treatment Effect (CATE), which measures the effect of a treatment on an outcome, conditional on covariates \citep{curth2021nonparametric, kennedy2023optimal, lipkovich2024modern}. In this section, we present one approach to using the meta-learner framework in the context of controlled subgroup selection. 

First, we introduce individual treatment effects (ITE) and CATE formally. We will work in the setting of potential outcomes \citep{hernan2025causal}; that is, we observe a triple $(Y_i, T_i,X_i)$ for each patient $i \in \{1,\ldots,n\}$, where $Y_i$ is a real-valued response, $T_i$ a binary treatment indicator taking values in $\{0,1\}$ and $X_i$ a $d$-dimensional, real covariate vector. Throughout, we assume these triples are independent and identically distributed between patients.  In many settings, $T_i = 0$ corresponds to patient $i$ having been assigned to a control arm, while $T_i = 1$ signifies that they were assigned to a (novel) treatment. The \emph{potential outcomes} framework then builds on the idea that we observe $Y_i = Y_i(1)$ when $T_i = 1$ and $Y_i = Y_i(0)$ when $T_i = 0$, but we never have access to the responses under treatment and control, $Y_i(1)$ and $Y_i(0)$ respectively, for the same patient at the same time. The individual treatment effect is the unobservable difference between these two quantities;
\[
\mathrm{ITE}_i := Y_i(1) - Y_i(0).
\]
The conditional expectation of $\mathrm{ITE}_i$ given $X_i = x$ is what is known as the CATE at $x$. Under the assumption that $\P(T_i=t\mid X_i=x) > 0$ for all $x \in \R^d$ and $t \in \{0,1\}$, we may write the CATE as 
\[
\mathrm{CATE}(x):=\E\bigl(Y_i(1) - Y_i(0)\bigm| X_i=x\bigr) = \E\bigl(Y_i\bigm| T_i=1, X_i = x\bigr) - \E\bigl(Y_i\bigm| T_i=0, X_i = x\bigr).
\]
The task of subgroup selection for heterogeneous treatment effects is then to identify the largest set $S_{\tau}$ for which $\mathrm{CATE}(x) \geq \tau$ for all $x \in S_{\tau}$. 

Now, if in addition to the observed response $Y_i(T_i)$, we would observe the counterfactual response $Y_i(1-T_i)$, we could take $Y_i(1)-Y_i(0)$ as the input for the response in the methods described in the previous subsections, as the covariate-conditional expectation of that difference is exactly the CATE, so that the output of the above methods would give an estimate of $S_{\tau}$. Unfortunately, we do not observe the counterfactual response, and this is where the meta-learning framework may be useful.

A range of meta-learning techniques have been discussed in the literature (see for instance the review by \cite{lipkovich2024modern}) and we will be focusing on using the so-called doubly robust (DR) learner for generating pseudo-observations mimicking the ITEs \citep{kennedy2023optimal}. The idea here is to use flexible machine learning methods in order to estimate the propensity $\pi: x \mapsto \P(T_i =1\mid X_i = x)$ and the treatment-conditional regression functions $\regressionFunction_t: x \mapsto \E\bigl\{Y_i(T_i)\mid T_i = t, X_i = x\bigr\}$ for $t \in \{0,1\}$. Writing $\hat{\pi}$ and $\hat{\eta}_t$ for estimators of these quantities, we then generate pseudo-observations as
\[
    \Tilde{Y}_i := \hat{\regressionFunction}_1(X_i) - \hat{\regressionFunction}_0(X_i) + \frac{T_i - \hat{\pi}(X_i)}{\hat{\pi}(X_i)\bigl(1 - \hat{\pi}(X_i)\bigr)}\bigl(Y_i - \hat{\regressionFunction}_{T_i}(X_i)\bigr)
\]
for $i\in \{1, \ldots, n\}$ using sample splitting (as will be described in more detail in Section~\ref{sec:simulations_HTE_methods}). The construction of the pseudo-observations $\Tilde{Y}_i$, $i \in \{1,\ldots,n\}$, as proxies for the individual treatment effects $\mathrm{ITE}_i$, $i \in \{1,\ldots,n\}$, in this way is rooted in the theory of efficient influence functions for estimating the CATE when neither the outcome functions $\eta_t$, $t \in \{0,1\}$, nor the propensity $\pi$ are known \citep{hines2022demystifying}. Indeed, even though we are interested in the analysis of randomised control trials, where the propensity is known, we nonetheless focus on the DR learner due to its robustness against misspecification of the outcome model as observed by \cite{sechidis2025usingindividualizedtreatmenteffects}. 

With these pseudo-observations in place, we may apply the methods described in the previous subsections to the data $(X_1, \tilde{Y}_1),\ldots,(X_n,\tilde{Y}_n)$. While this approach allows for a high degree of flexibility and to leverage the predictive capabilities of modern machine learning tools, finite-sample guarantees are difficult to achieve. Section~\ref{sec:simulations_HTE} below investigates the performance empirically in a simulation study, where the data-generating distributions are chosen to closely match those observed for real-world clinical data using software by \cite{sun2024benchtm}.

\section{Simulation study}
\label{sec:simulation_study}

The work by \cite{wan2024confidence} and \cite{mueller2024isotonic} contains proofs of the Type I error guarantee under suitable assumptions for the main methods presented in Section~\ref{sec:methods}. The aim of this simulation study is to identify how violations of these assumptions affect the ability of the methods to correctly identify subgroups while controlling Type I error rates. More specifically, we are interested in clinical trial settings like those presented in Section~\ref{sec:application_clinical_trials}. While in such settings one may have some domain knowledge of the data-generating mechanism, one will sometimes not be able to say with certainty that the assumptions of either presented method are fully satisfied. Hence, before we can use our methods to identify subgroups of high risk of adverse events (AEs) (Section~\ref{sec:application_clinical_trials_AE})  or high treatment effect with our study data (Section~\ref{sec:application_clinical_trials_TEH}), we need to understand the reliability of these results when the true mechanisms do not align with the assumptions needed to guarantee Type I error rate control. Moreover, we also seek to quantify the effectiveness of the discussed methods at selecting subgroups of interest.

While it is fundamentally impossible to comment on the relationship between the selected subgroups and the unknown true subgroups in Section~\ref{sec:application_clinical_trials}, and hence to determine whether a Type I error has been committed or not, say, through our simulation study, we will be able to answer questions such as:\vspace{0.5em}
\begin{adjustwidth}{2em}{2em} 
    \textit{``If the true relationship was non-monotone (and hence non-linear), how would that affect the Type I error control of the two methods?''},
\end{adjustwidth}
or
\begin{adjustwidth}{2em}{0em} 
    \textit{``If the true relationship were correctly described by a logistic model, what would the power of the two methods be?''}    
\end{adjustwidth}
\vspace{0.3em}which provides important insights into the behaviour of the considered methods.

We now first present the simulation study for the application of identifying subgroups at high risk of AEs, before continuing to the problem of treatment effect heterogeneity (TEH). For both studies, the presentation of setup and results will follow the ADEMP (Aims, Data, Estimand, Methods and Population) framework \citep{morris2019using}.

\subsection{Discovering the subgroup at low risk of adverse effects}\label{sec:simulations_AE}

We start our simulation study with the setting of Section~\ref{sec:application_clinical_trials_AE}, where we let $Y_i$ be equal to $1$ if no AE occurred for the $i$th patient and $0$ otherwise. Hence, $\regressionFunction(x)$ gives the probability of patient $i$ not facing an adverse event if $X_i = x$.

\subsubsection{Data generation}\label{sec:simulations_AE_}

\paragraph{Choice of covariates}
Throughout, we include a proxy for a patient's drug exposure in $X,$ for example this can be the highest drug concentration in the blood after dosing. In Section~\ref{sec:simulations_AE_univariate} this will be the only covariate. In the subsequent Sections~\ref{sec:simulations_AE_categorical} and~\ref{sec:simulations_AE_multivariate}, we then look at how the inclusion of continuous and categorical covariates affects the methods' behaviours.
Some variables have been non-linearly transformed prior to the analysis and all have been rescaled to take values in $[0,1]$, with the binary variable being represented by a variable taking values in $\{0,1\}$.

\paragraph{Ground truths}
For each collection of covariates we then define several functions $\regressionFunction$ that will serve as potential ground truth regression functions. These were selected so that important cases of deviation, either slight or severe, from the monotonicity and linearity assumptions on which the Type I error rate guarantees of the investigated methods respectively rely, are considered. Details are given in Sections~\ref{sec:simulations_AE_univariate},~\ref{sec:simulations_AE_categorical} and~\ref{sec:simulations_AE_multivariate} respectively for the three considered scenarios of included covariates. It is worth pointing out that these ground truth regression functions were generated by using a regression technique on the original study data and then transforming the resulting function to either yield sufficient signal for a meaningful comparison of methods (e.g.~through rescaling) and/or to exhibit certain behaviour of interest (e.g.~``Logistic model (truncated)'' as described in Section~\ref{sec:simulations_AE_univ_setup_and_methods}).

\paragraph{Sampling pseudo-random covariate-response-pairs}

We refer to a \emph{simulation scenario} as the combination of a set of covariates and a ground truth regression function. For each of these simulation scenarios, we simulate datasets of sizes $n \in \{125, 250, 500, 1000\}$, which we hope to cover a large range of the sample sizes encountered in clinical applications. For each fixed $n$, we then follow the following two-step procedure.

Firstly, we sample covariate values that mimic the covariate distribution of the original case study in Section~\ref{sec:application_clinical_trials_AE} by using the package \texttt{syn}-function from the \texttt{R}-package \pkg{synthpop} \citep{nowok2016synthpop} with default parameters.
Secondly, at each sampled covariate point $X_i$, we independently sample the response $Y_i$ from a Bernoulli distribution with success probability $\eta(X_i)$ for the chosen ground truth $\regressionFunction$.

We repeat the process described above independently $B = 200$ times, resulting in $B$ datasets for each combination of a simulation scenario with $n$.

\subsubsection{Estimand: the target subgroup}\label{sec:simulations_AE_target}
The notion of a subgroup we consider in this paper is defined as $\superlevelSet = \{x: \regressionFunction(x) \geq \threshold\}$, for some threshold $\tau \in \R$ chosen by the practitioner, making $\superlevelSet$ the estimand of interest. In most simulation scenarios, we choose $\tau$ such that roughly 50\% of the patient population should be expected to fall into $\superlevelSet$.
Exceptions to this choice of $\threshold$ exist and details are provided in Section~\ref{sec:simulations_AE_univariate}, \ref{sec:simulations_AE_categorical} and \ref{sec:simulations_AE_multivariate}.

\subsubsection{Methods}
In this simulation study, we compare the methods described in Sections~\ref{sec:methods_glm} and~\ref{sec:methods_iss}. Both methods take as input the threshold $\threshold$, a nominal Type I error rate, which we take to be $\alpha = 0.1$ throughout, and a dataset consisting of covariate-response pairs. They then each output a subset of the covariate domain. 

For the GLM-based method in Section~\ref{sec:methods_glm}, one needs to specify a GLM (i.e.~which covariates to include, whether to include interaction effects, a link function, etc.\footnote{In each application of the GLM-based method, the critical values $\hat{c}_1$ and $\hat{c}_2$ appearing in Section~\ref{sec:methods_glm} have to be found via simulation, whose number of replications we also need to specify. Throughout the simulation study at hand, we used $1000$ replications per application of the method. We refer the reader to \cite{wan2024confidence} and references therein for details on how $\hat{c}_1$ and $\hat{c}_2$ are found.}). Throughout, we use all covariates as predictors and do not include any interaction effects. Moreover, noting that we are in a classification setting as our responses are binary, we use a logit link function. 

For the ISS method, we do not need to specify any sort of parametric model, and instead only need to specify for each covariate, whether an increase in it will lead to an increase or decrease in $\regressionFunction$ (across the entire covariate domain). We will assume that exposure to the compound is positively associated with facing AEs. Hence, since $\regressionFunction$ gives the conditional probability of staying free of AEs, ISS will be applied under the assumption that $\regressionFunction$ is decreasing in this variable.

\subsubsection{Performance metrics}\label{sec:simulations_AE_metrics}
We evaluate the empirical properties of the subgroup selection methods by analysing the selected sets resulting from application of our methods to the $B$ independently generated data sets for each simulation scenario and sample size $n$. To analyse the set selected by a method at the $b$th independent replication (for $b \in \{1,\ldots, B\}$), we use once more the \texttt{R}-package \textbf{synthpop} to draw from the patient population a Monte Carlo (MC) sample of $M = 10^5$ independent covariate observations $X^{(b, 1)}, \ldots, X^{(b, M)}$, based on which we then evaluate the set's properties.

The specific properties we are interested in can be grouped into either characterising the \emph{reliability} (through Type I error rate and a notion we call \emph{false selection rate}) or the \emph{effectiveness} of the selected procedure for selecting the subgroup. We present this discussion for the case of lower bounding $\superlevelSet$ only, i.e.~when evaluating $\setLowerBound$. Corresponding metrics for $\setUpperBound$, $\setLowerBoundTwoSided$ and $\setUpperBoundTwoSided$ could also be defined. See Table~\ref{table:measures_of_interest_lowerbound} for an overview over the metrics of interest, which we now describe in more detail. 

\paragraph{Strict reliability: Type I error rate control}

The Type I error rate gives the probability of $\setLowerBound$ selecting points that do not fall into $\superlevelSet$. We estimate this via the proportion of the $B$ independent replications for which at least one of the $M$ MC points falls into $\setLowerBound$ but not $\superlevelSet$. 

\paragraph{Weak reliability: False selection rate}
The false selection rate (FSR) is a more relaxed measure than the Type I error rate, and is given by the average proportion of~$\setLowerBound$ not falling into $\superlevelSet$. More specifically, this proportion is given by the probability of a new patient sampled from the patient population restricted to $\setLowerBound$, not actually belonging to the true subgroup $\superlevelSet$; i.e.~$\P(X_0 \in \setLowerBound\setminus\superlevelSet \mid X_0 \in \setLowerBound)$, 
which captures the randomness in both $X_0$ and the training data used to fit $\setLowerBound$.
While the Type I error rate tells us the proportion of cases in which any part of $\setLowerBound$ is not contained in $\superlevelSet$,
the false selection rate tells us what fraction of patients classified as belonging to the subgroup should not have been categorised as such. 
To illustrate the difference between the false selection rate and the Type~I error rate for the task of lower bounding $\superlevelSet$, suppose we have a method whose Type I error rate is 90\% and whose false selection rate is 1\%. While this method classifies \emph{some} patients who are not part of the subgroup as belonging to it in 9 out of 10 cases, only 1 out of 100 patients from the training population are on average affected by such wrong categorisations.
Indeed, in certain applications in which reliability is less critical, a large Type I error rate might not be such an issue, as long as the false selection rate is very small.  It is also reasonable to consider these quantities in relation to one another; the false selection rate can never exceed the Type I error rate, but values close to it indicate high severity of Type I errors when they occur.
We finally remark that while the Type I error rate is invariant under covariate shifts between the training population used to fit $\setLowerBound$
and the population of patients that should be categorised as part of the subgroup or not, the false selection rate is not and may increase dramatically if the distribution from which the new patient $X_0$ is sampled changes. For instance, while the set $\hat{L}_\alpha\setminus S_\tau$ may be small in terms of the original distribution of patients, a covariate shift could lead to this set containing a new patient $X_0$ with high probability. 

\paragraph{Effectiveness: Power}
The most natural measure for the \emph{effectiveness} of the procedure, is the similarity of $\setLowerBound$ to $\superlevelSet$. More specifically, an effective procedure will have high power in the sense that $\superlevelSet\setminus\setLowerBound$ is a small set on average, or equivalently $\superlevelSet\cap\setLowerBound$ is close to $\superlevelSet$. Weighting by the covariate distribution, we define the \emph{power} as $\P(X_0 \in \setLowerBound \mid X_0 \in \superlevelSet) = \P(X_0 \in \setLowerBound \cap \superlevelSet)/\P(X_0 \in \superlevelSet)$.

\begin{table}[tb]
    \resizebox{\textwidth}{!}{
    \setlength{\tabcolsep}{0.5em} 
    \renewcommand{\arraystretch}{1.5}
    \centering
    \begin{tabular}{c|c|c}
          \toprule Property & Definition & Estimator \\
         \toprule
         Type I error rate & $\P\bigl(\setLowerBound \nsubseteq \superlevelSet\bigr)$ & $\frac{1}{B}\sum_{b=1}^B \one \bigl\{\exists m \in \{1,\ldots,M\}: X^{(b,m)} \in \setLowerBound^b \setminus \superlevelSet\bigr\}$ \\
         False selection rate & $\P\bigl(X_0 \in \setLowerBound\setminus\superlevelSet \bigm| X_0 \in \setLowerBound\bigr)$ & $\frac{1}{B}\sum_{b=1}^B\Bigl(\sum_{m=1}^M \one \bigl\{X^{(b, m)} \in \setLowerBound^b\setminus\superlevelSet\bigr\}\bigm/\sum_{m=1}^M \one \bigl\{X^{(b, m)} \in \setLowerBound^b\bigr\}\Bigr)$ \\
         Power & $\P\bigl(X_0 \in \setLowerBound \bigm| X_0 \in \superlevelSet\bigr)$ & $\frac{1}{B}\sum_{b=1}^B\Bigl(\sum_{m=1}^M \one\bigl\{X^{(b,m)} \in \setLowerBound^b \cap \superlevelSetCustomArg{\threshold}\bigr\} \bigm/ \sum_{m = 1}^M \one\bigl\{X^{(b,m)} \in \superlevelSetCustomArg{\threshold}\bigr\}\Bigr)$\\
         \bottomrule 
    \end{tabular}}
    \caption{Performance measures for methods outputting $\hat{L}_\alpha$ with the aim of lower bounding $S_\tau$. For the purpose of evaluating the estimators, we set $0/0 := 0$.}
    \label{table:measures_of_interest_lowerbound}
\end{table}

\subsubsection{Univariate subgroup selection}\label{sec:simulations_AE_univariate}

We start these simulations by considering a univariate setting, in which we are interested to identify for which levels of our proxy of drug exposure $X$, the probability $\eta(X)$ of staying free of adverse events (AEs) is above a pre-defined threshold. 

\paragraph{Setup \& Methods}\label{sec:simulations_AE_univ_setup_and_methods}
We consider five ground truth functions $\eta$ in this simulation, all of which are (at least roughly) decreasing, to capture that lower exposure should lead to a lower chance of AEs. To the data generated from these ground truths, we then apply the two methods described above; the GLM-based method with a logistic link and ISS with the assumption of a decreasing relationship.

We have summarised the considered settings in Table~\ref{table:simulations_AE_univariate_table_of_functions}, which also shows which methods are correctly specified for which ground truth, and hence their theoretical suitability to control the Type I error rate. It is worth highlighting that the function labelled ``Logistic model (truncated)'' has $\superlevelSet = \emptyset$.  Here, we took the ground truth regression function and $\tau$ of the case ``Logistic model'' and constructed $\regressionFunction$ to be the pointwise maximum value between the ground truth regression function of the logistic model and $\tau - 0.01$. Hence, $\regressionFunction$ will here look like the logistic function in ``Logistic model'', but capped (or truncated) just below the threshold $\threshold$. We include this case as a negative control, to assess the procedures' ability to select the empty set whenever there is no true subgroup to select.

\begin{table}[tbp]
    \resizebox{\textwidth}{!}{
    \setlength{\tabcolsep}{0.5em} 
    \renewcommand{\arraystretch}{1.35}
    \centering
    \begin{tabular}{c|c|c c |c c} \toprule
        Ground truth $\eta$ & Figure & $\tau$ & Size of $\superlevelSet$ & GLM (logistic) & ISS (decreasing)\\ \toprule
        Logistic model & Fig.~\ref{fig:AE_logmod_results} & 0.76 & 0.50& \checkmark & \checkmark \\ \hline
        Logistic model (truncated)& \multirow{ 2}{*}{Fig.~\ref{fig:AE_partially_misspecified}} & 0.76 & 0  & & \checkmark \\ 
        Step function & & 0.34 & 0.78 &  & \checkmark \\ \hline
        Logistic model (quadratic component) & \multirow{ 2}{*}{Fig.~\ref{fig:AE_fully_misspecified}} & 0.75 & 0.50 & & \\
        Non-smooth \& non-monotone & & 0.38 & 0.55 & & \\ \bottomrule 
    \end{tabular}}
    \caption{Each row gives the label of a considered ground truth, which is illustrated in the first panel of the associated figure. We also provide the chosen threshold $\tau$, the size of the subgroup $S_\tau$ (in terms of the proportion of the covariate distribution falling into $S_\tau$) rounded to two digits, and whether each considered method is correctly specified, in which case a tick mark is present. Based on equal patterns of correct specification, the rows are grouped into three sections.}
    \label{table:simulations_AE_univariate_table_of_functions}
\end{table}

\begin{figure}[t]
    \begin{subfigure}[c]{0.24\textwidth}
        \centering
        \includegraphics[width=\textwidth]{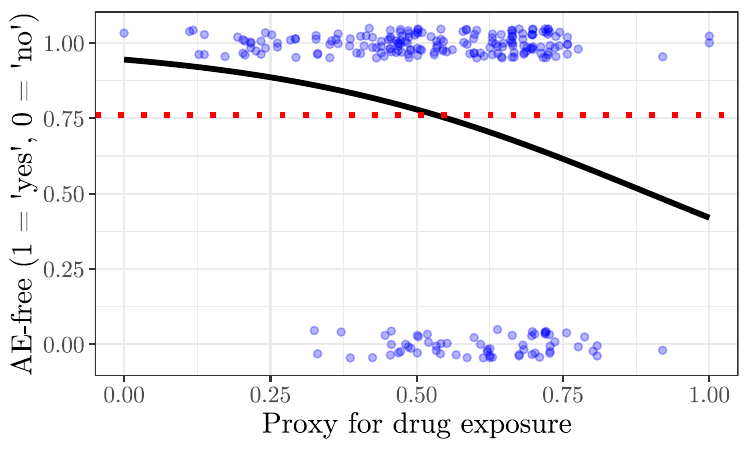}
    \end{subfigure}
    \begin{subfigure}[c]{.75\textwidth}
        \centering
        \includegraphics[width=\textwidth]{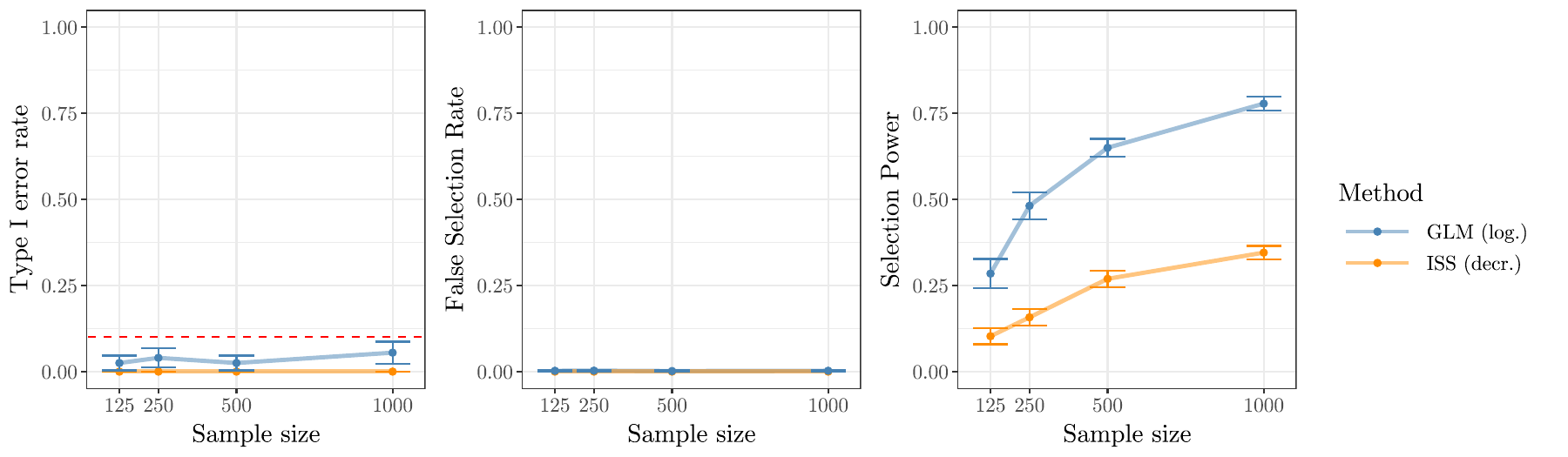}
    \end{subfigure}
    \caption{Data generation and empirical performance for the ground truth ``Logistic model''. The leftmost panel shows the true regression function (solid, black line), the threshold $\tau$ (dotted, red line) and one out of $B$ independent draws of $n = 250$ data points (blue dots). The other panels illustrate, in order, the estimated Type I error rate, false selection rate and power. The red dashed line in the first of these panels gives the chosen nominal Type I error rate $\alpha = 0.1$. In all three figures, the $x$-axes gives the sample size $n$. Error bars provide approximate $95$\% confidence intervals based on a normal approximation of the distribution of the mean of the $B$ independent replications.}
    \label{fig:AE_logmod_results}
\end{figure}

\paragraph{Results: Warm-up --- the logistic model as ground truth}
We start with the most benign case, where the ground truth corresponds to a logistic model (see the leftmost panel in Figure~\ref{fig:AE_logmod_results}). Here, the GLM-based method and ISS are correctly specified and as can be seen in the second panel of the figure, both methods indeed control the Type I error rate, although the GLM-based method is closer to the nominal value of $\alpha = 0.1$. This naturally also results in a small FSR for both methods. Unsurprisingly, the GLM-based method, which is specifically tailored to the functional form of the population regression function considered here, results in a higher effectiveness, as depicted in the rightmost panel.

\begin{figure}[tbp]
     \centering
     \begin{subfigure}[c]{0.24\textwidth}
        \centering
        \includegraphics[width=\textwidth]{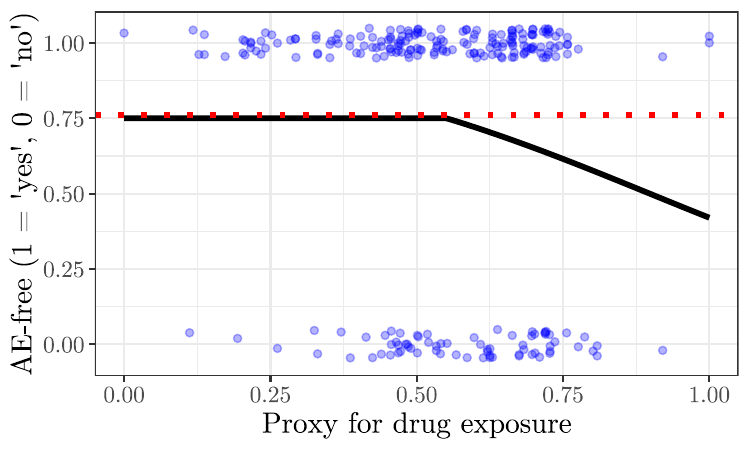}
    \end{subfigure}
    \begin{subfigure}[c]{.75\textwidth}
        \centering
        \includegraphics[width=\textwidth]{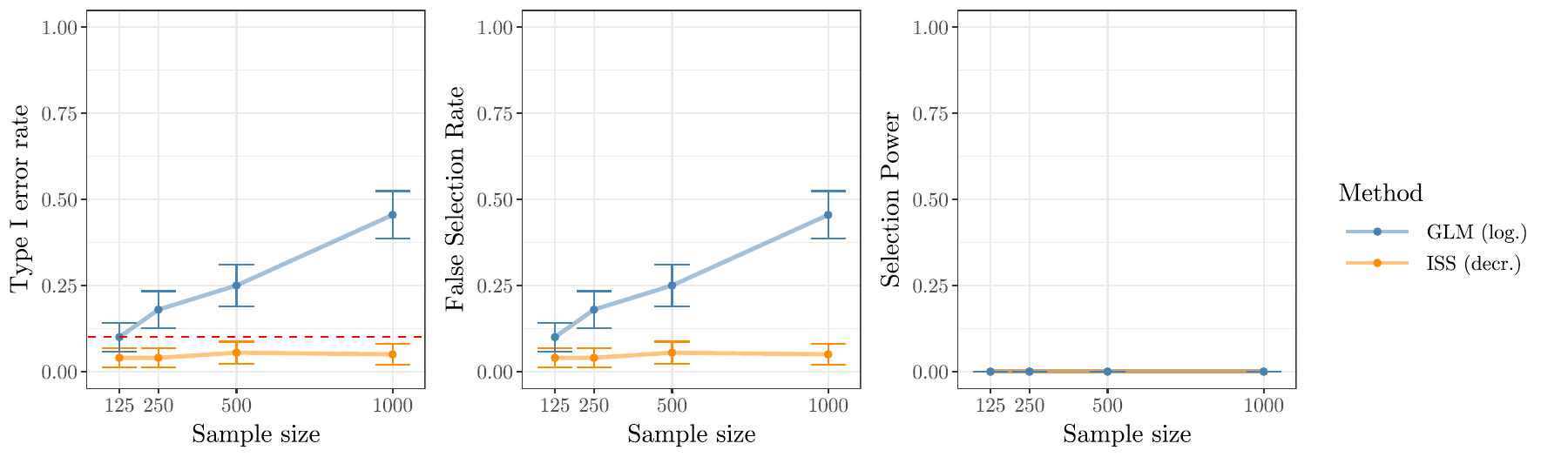}
    \end{subfigure}%
     \\ \vspace{0.2cm}
     \begin{subfigure}[c]{0.24\textwidth}
        \centering
        \includegraphics[width=\textwidth]{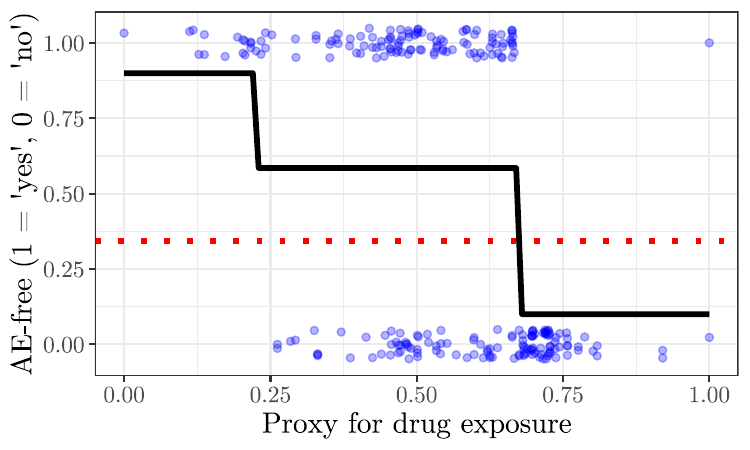}
    \end{subfigure}
    \begin{subfigure}[c]{.75\textwidth}
        \centering
        \includegraphics[width=\textwidth]{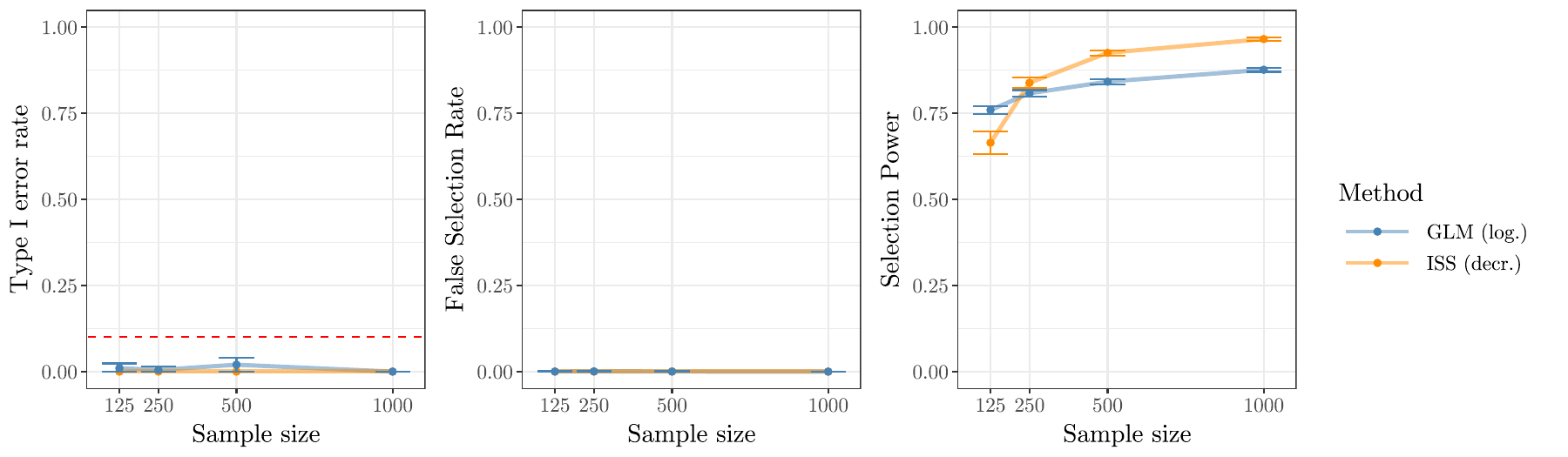}
    \end{subfigure}
    \caption{Data generation and empirical performance for the ground truths ``Logistic model (truncated)'' (top row) and ``Step function'' (bottom row). See caption of Fig.~\ref{fig:AE_logmod_results} for further details.}
        \label{fig:AE_partially_misspecified}
\end{figure}

\paragraph{Results: Deviations from linearity}
Figure~\ref{fig:AE_partially_misspecified} presents results for cases in which the GLM-based method is no longer correctly specified, but ISS is, since $\eta$ is still monotonically decreasing. \redCommentToBeRemoved{(This line break to be removed.)}In the setting labelled ``Logistic model (truncated)'', the true subgroup $S_\tau$ is empty, and any Type~I error immediately renders the entire selected set to be falsely selected, so that Type I error rate and false selection rate coincide. Moreover, the true subgroup being empty also renders the question of a method's power nonsensical.

Since the GLM-based method supposes a linear model and hence a ``global'' structure, it is unable to capture the truncation in ``Logistic model (truncated)'', leading to Type I error rates above the nominal level $\alpha$. \redCommentToBeRemoved{(This line break to be removed.)}However, in ``Step function'', we see a case in which the deviation from linearity does not lead to an inflated Type~I error rate of the GLM-based procedure. Since both settings concern decreasing regression functions, we have that ISS controls the Type I error rate, as guaranteed by the associated theory. Concerning the efficacy in ``Step function'', we see that both methods achieve high power.

\begin{figure}[tbp]
     \centering
     \begin{subfigure}[c]{0.24\textwidth}
        \centering
        \includegraphics[width=\textwidth]{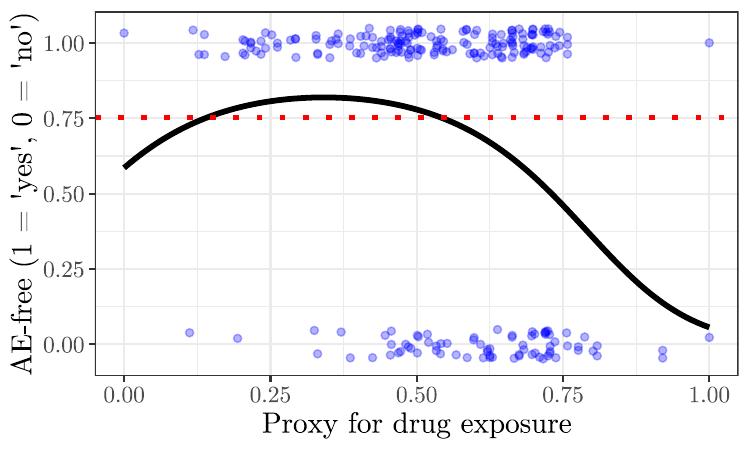}
    \end{subfigure}
    \begin{subfigure}[c]{.75\textwidth}
        \centering
        \includegraphics[width=\textwidth]{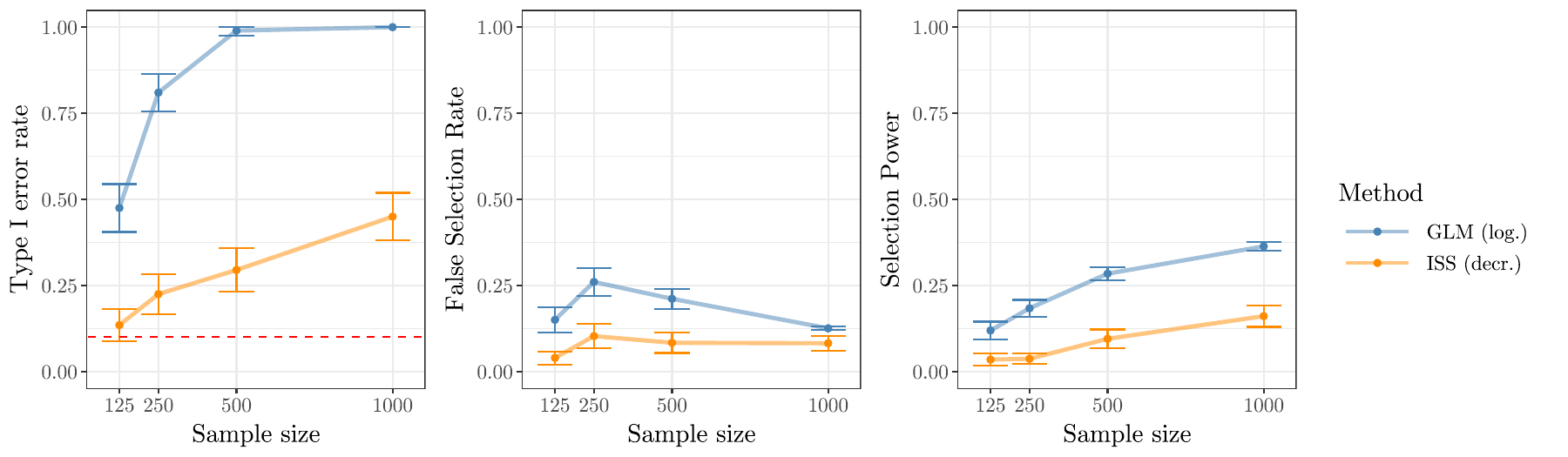}
    \end{subfigure}%
     \\ \vspace{0.2cm}
     \begin{subfigure}[c]{0.24\textwidth}
        \centering
        \includegraphics[width=\textwidth]{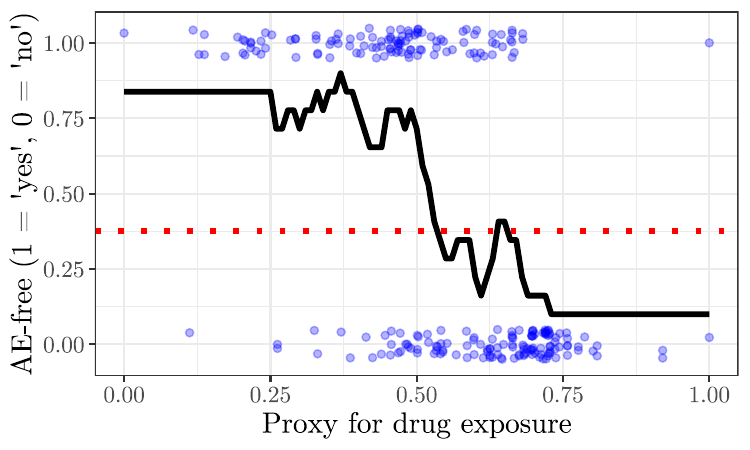}
    \end{subfigure}
    \begin{subfigure}[c]{.75\textwidth}
        \centering
        \includegraphics[width=\textwidth]{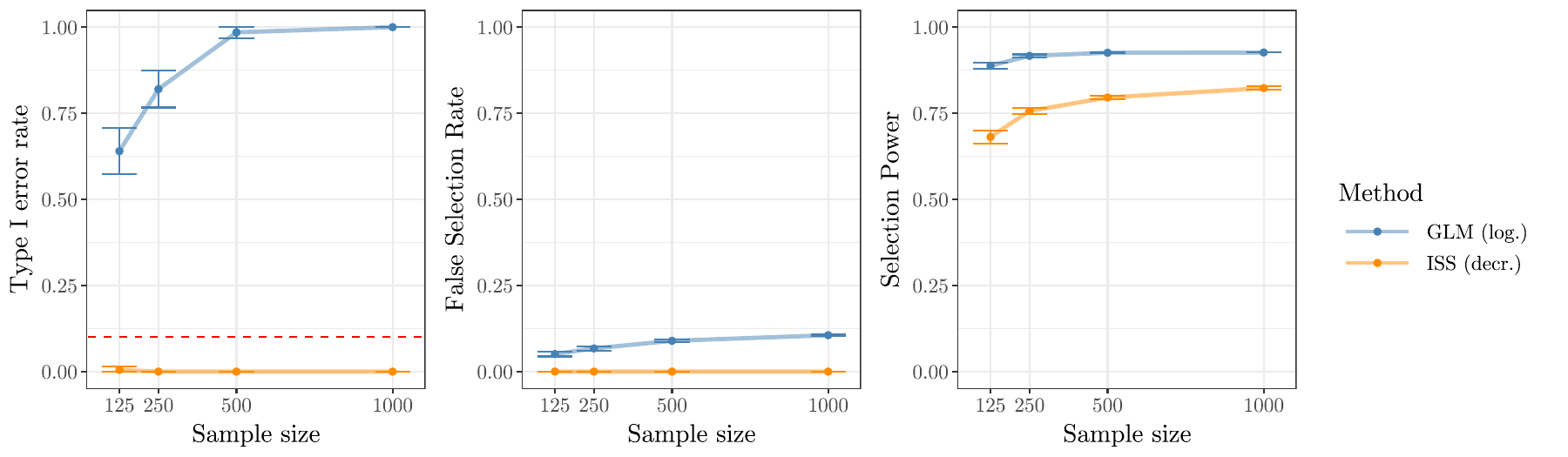}
    \end{subfigure}
        \caption{Data generation and empirical performance for the ground truths ``Logistic model (quadratic component)'' (top row) and ``Non-smooth \& non-monotone'' (bottom row). See caption of Fig.~\ref{fig:AE_logmod_results} for further details.}
        \label{fig:AE_fully_misspecified}
\end{figure}

\paragraph{Results: Deviations from monotonicity}\label{sec:simulations_AE_univ_deviations_from_monotonicity}
In Figure~\ref{fig:AE_fully_misspecified}, we have two settings for which no considered method is correctly specified. However, in the case ``Logistic model (quadratic component)'', we have a logistic regression model with a quadratic component and if the practitioner knew about the existence of such a quadratic component, they could use the GLM-based method and account for it in the model formulation. Here, we consider the setting in which the practitioner fails to include a quadratic term. We see that here both the GLM-based method and ISS supposing a decreasing relationship far exceed the nominal Type I error rate. 

For the case of the ``Non-smooth \& non-monotone'' regression function on the other hand, it is worth mentioning that ISS has a robustness property as a result of which it is  not necessary to know the direction of the relationship exactly. Indeed, when applying ISS with a decreasing relationship in mind, it is sufficient for Type I error control that --- within the span of of the covariate distribution --- the true subgroup is an interval that is unbounded to the left. Therefore, the regression function may oscillate without violating the Type I error guarantee of ISS, as long as this oscillation does not lead to any value to the left of a value in the subgroup to no longer be in the subgroup (we shall return to this point below). While this is not exactly satisfied here, it seems we are close enough to still have Type I error rate control. In comparison, the GLM-based method has a Type~I error rate that is close to $1$ for large sample sizes, though the false selection rate remains very small. 

While the last column of Figure~\ref{fig:AE_fully_misspecified} illustrates the observed power of the considered procedures, it is important to not overstate their importance: for $3$ out of $4$ of the applications of subgroup selection methods here (i.e.~all except applying ISS to ``Non-smooth \& non-monotone''), the Type I error rate is not controlled, rendering efficacy considerations irrelevant. After all, reliability is the primary goal in subgroup selection, and if we were interested in power in cases where Type I error rate control is not given, we could ignore all data and simply output the entire covariate domain. Of course, this would not be a reasonable approach in practice and the strong reliability guarantees are the key benefit of our framework and the methods we consider (under appropriate assumptions violated here for illustration purposes).

\paragraph{Robustness of ISS}\label{sec:simulations_AE_univ_ISS_robustness}
In Figure~\ref{fig:AE_fully_misspecified}, we have seen that in the setting labelled ``Non-smooth \& non-monotone'', ISS still controls the Type I error rate even though monotonicity is not given. As explained, this is driven by a robustness property of ISS, allowing it to retain validity as long as the true subgroup is an interval\footnote{In multivariate settings, an analogous robustness property holds, whereby --- under the assumption of a decreasing regression function in all covariates --- $x \in S_\tau \subseteq \R^d$ needs to imply $x' \in S_\tau$ if $x'$ is smaller than $x$ in every coordinate. In other words, $S_\tau$ needs to be a so-called \emph{lower set}.} that is unbounded to the left. The effect driving this in univariate settings is that ISS works by testing inclusion in the true subgroup for many points in the covariate domain, before accounting for the multiplicity of tests via a fixed-sequence procedure, as described in Section~\ref{sec:methods}. Since each of these tests only assesses whether or not the point at which we test is in the true subgroup, monotonicity only comes in to determine the order of the sequence of tests. At the same time, the specific test statistic used does not rely on the monotonicity either. 

Of course, this does not make ISS a panacea, and deviation from monotonicity can be severe enough that even this weaker requirement is no longer satisfied, as illustrated in the case ``Logistic model (quadratic component)''. Since that case would correspond to a lower drug exposure leading to a lower probability of staying free of AEs, this is not a realistic concern in practice, however, and is only included for illustrative purposes of behaviour when the validity breaks down.

\paragraph{Conclusions}
The simulations in these univariate scenarios give an idea of the performance and sensitivity to assumptions of the considered methods. Throughout, we see that the presented methods are effective tools for conducting subgroup selections and confirm the control of the Type I error rate under correct specification. While ISS is able to control Type I error rate over an even larger family of ground truth regression functions, it may lead to smaller selected sets, as seen for instance in Figure~\ref{fig:AE_logmod_results}, where the GLM is correctly specified and therefore exhibits excellent efficacy and reliability at the same time. 

In the next sections, we consider in turn what happens when including additional continuous and categorical covariates.

\subsubsection{Multivariate subgroup selection with a categorical variable}\label{sec:simulations_AE_categorical}
We now extend the previous section's setting by also including a binary variable. Our focus here is to examine the effect of directional specification of the binary variable in ISS, and how the antichaining method described in Section~\ref{sec:methods_iss_categorical} may help when the direction is not known for certain.

\begin{figure}[ht]
     \centering
     \begin{subfigure}[c]{0.24\textwidth}
        \centering
        \includegraphics[width=\textwidth]{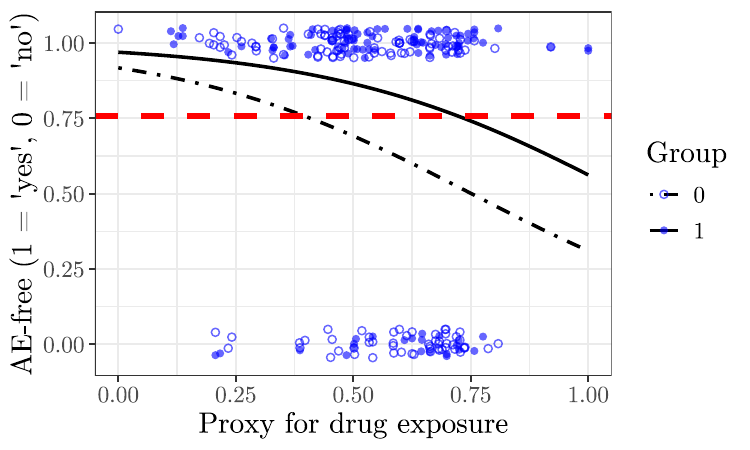}
    \end{subfigure}
    \begin{subfigure}[c]{.75\textwidth}
        \centering
        \includegraphics[width=\textwidth]{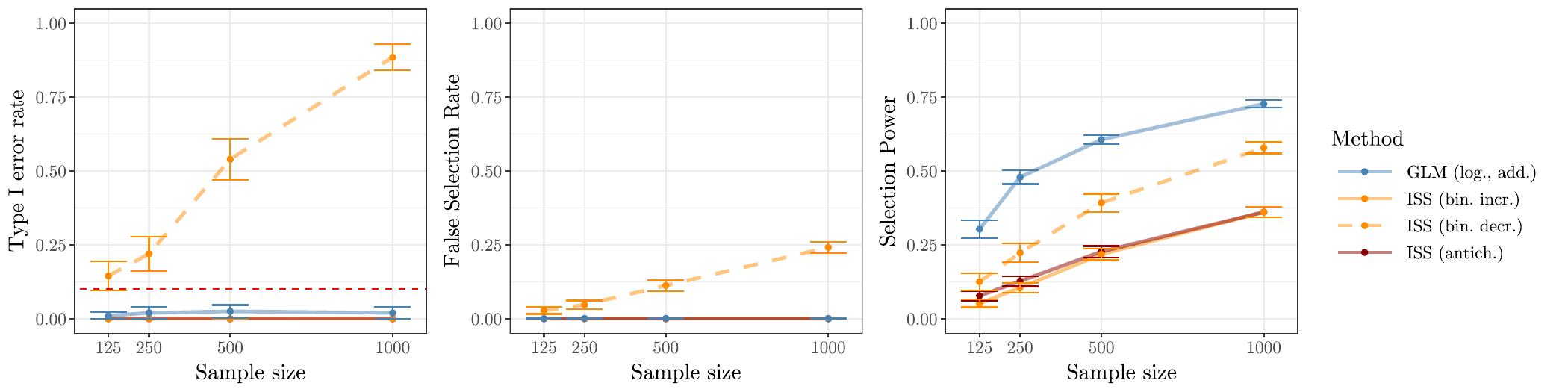}
    \end{subfigure}%
    \\ \vspace{0.2cm}
     \begin{subfigure}[c]{0.24\textwidth}
        \centering
        \includegraphics[width=\textwidth]{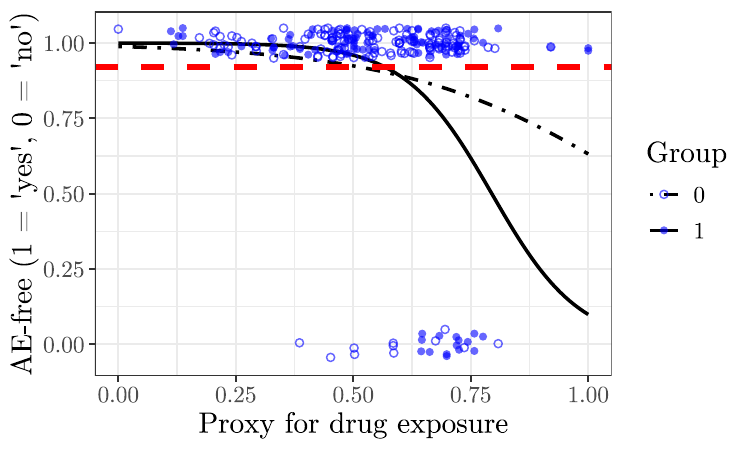}
    \end{subfigure}
    \begin{subfigure}[c]{.75\textwidth}
        \centering
        \includegraphics[width=\textwidth]{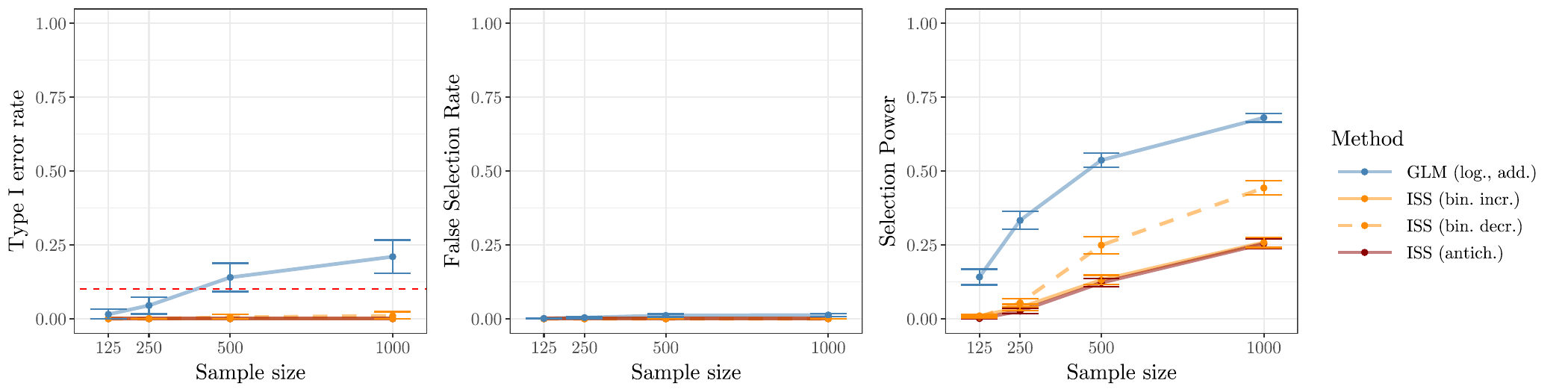}
    \end{subfigure}%
     \\ \vspace{0.2cm}
     \begin{subfigure}[c]{0.24\textwidth}
        \centering
        \includegraphics[width=\textwidth]{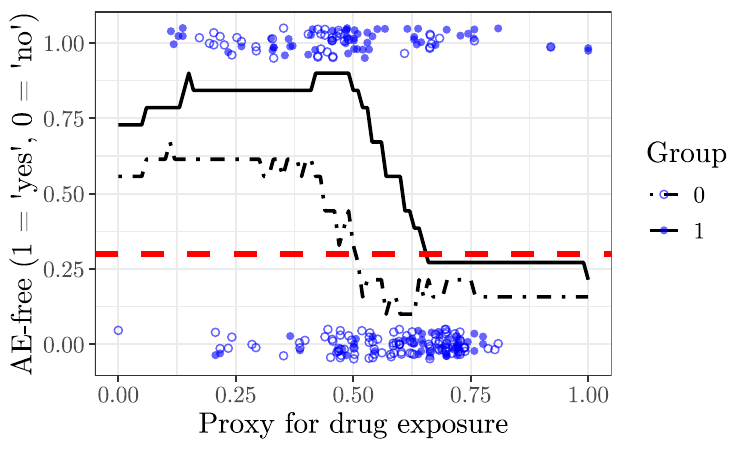}
    \end{subfigure}
    \begin{subfigure}[c]{.75\textwidth}
        \centering
        \includegraphics[width=\textwidth]{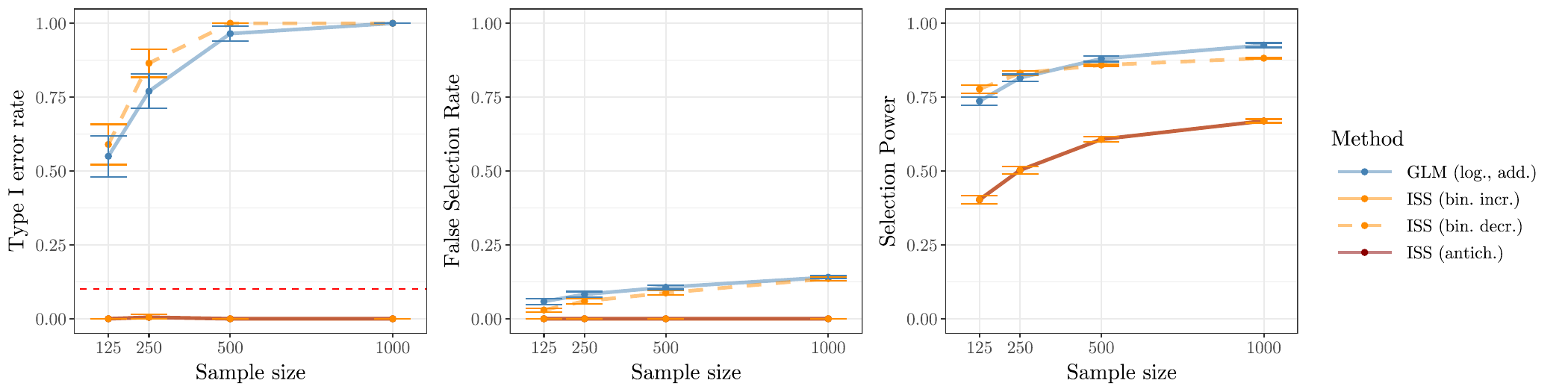}
    \end{subfigure}
        \caption{Data generation and empirical performance for the ground truths ``Logistic model (additive)'' (top row), ``Logistic model (interaction)'' (middle row) and ``Non-smooth \& non-monotone'' (bottom row). The leftmost column illustrates the respective regression functions, where the linetype represents the value of the binary variable (labelled ``Group'' in the legend) and the red dotted line gives the value of $\tau$. See the caption of Fig.~\ref{fig:AE_logmod_results} and the text for details on the other three columns.}
        \label{fig:AE_biv_with_categorical}
\end{figure}

\paragraph{Setup \& Methods}
We here consider three ground truth regression functions, which are visualised in the first column of Figure~\ref{fig:AE_biv_with_categorical} and listed in Table~\ref{table:simulations_AE_binary_table_of_functions}. ``Logistic model (additive)'' and ``Logistic model (interaction)'' differ in whether or not an interaction effect between the binary variable and the drug exposure has been included in the data-generating distribution. We then apply four variants of the considered methods. For the GLM-based method, we again use a logistic link and do not include an interaction term in the model specification. However, we apply ISS now always under the assumption of a decreasing effect of drug exposure, but vary whether a change from Group 0 to 1 is envisioned to have a positive (``ISS (binary incr.)'') or negative effect (``ISS (binary decr.)''), or whether the approach in Section~\ref{sec:methods_iss_categorical} (``ISS (antichaining)'') is to be used. As above, before applying ISS we scale each covariate separately to have unit variance, which induces a distance between the two levels of the categorical variable. Unless the antichaining-technique is used, the covariate observations are then seen as elements of the two-dimensional Euclidean space, with the distance between points being influenced by said scaling.

Table~\ref{table:simulations_AE_binary_table_of_functions} illustrates the settings under which each method is correctly specified. 
We highlight that in the setting ``Logistic model (interaction)'', any value of the continuous covariate contained in the true subgroup when in Group 0 would also be in the true subgroup when switching to Group 1. Hence, by the robustness property of ISS highlighted in Sections~\ref{sec:simulations_AE_univ_deviations_from_monotonicity} and~\ref{sec:simulations_AE_univ_ISS_robustness}, we should still expect Type I error rate control through the method ``ISS (binary incr.)''.

\begin{table}[tbp]
    \resizebox{\textwidth}{!}{
    \setlength{\tabcolsep}{0.5em} 
    \renewcommand{\arraystretch}{1.35}
    \centering
    \begin{tabular}{c|c c |c c c c} \toprule
        Ground truth $\eta$ & $\tau$ & Size of $S_\tau$ & GLM (logistic, additive) & ISS (binary incr.) & ISS (binary decr.) & ISS (antichaining)\\ \toprule
        Logistic model (additive) & 0.76 & 0.50 & \checkmark & \checkmark &  & \checkmark\\ 
        Logistic model (interaction) & 0.92 & 0.50 & & & & \checkmark \\
        Non-smooth \& non-monotone & 0.30 & 0.56 & & & & \\ \bottomrule 
    \end{tabular}}
    \caption{See the caption of Table~\ref{table:simulations_AE_univariate_table_of_functions} and the text for details.}
    \label{table:simulations_AE_binary_table_of_functions}
\end{table}

\paragraph{Results}
We first consider the reliability of the discussed methods. For ISS, we can see that as expected through the robustness of the method to some violation of the monotonicity, the Type I error rate is controlled for all three ground truths, if the direction of the effect of the categorical variable is correctly specified. However, if the direction is misspecified, the Type I error is violated in all but the ground truth labelled ``Logistic model (interaction)''; here, while the regression function is very much non-constant in the binary variable, the boundary of the true subgroup is mostly constant. For the GLM-based method, it is particularly interesting to see that the omission of the interaction term (comparing the first and second row of the second column of Figure~\ref{fig:AE_biv_with_categorical}) is sufficient to lose control of Type I error rate control. When the method is correctly specified (i.e.~in the case ``Logistic model (additive)'' in the top row), the GLM-based method achieves excellent power. In all three ground truths, we see that even when Type I error rate control is violated, the false selection rates stay low. 

\paragraph{The antichaining approach in ISS}
In Section~\ref{sec:methods_iss_categorical}, we present the \emph{antichaining approach}. Figure~\ref{fig:AE_biv_with_categorical} illustrates that this approach can protect against the large Type~I error rates incurred due to directional misspecification of the binary variable, without significant deviation from the power achieved by an application of ISS with correct directional specification. Indeed, it can even lead to an increase in power, as the problem of scaling the binary variable appropriately disappears. As such, the antichaining method is a useful addition to the ISS method as presented by \cite{mueller2024isotonic} and extends its applicability even further.

\subsubsection{Multivariate subgroup selection} \label{sec:simulations_AE_multivariate}

All considered methods can be used for an arbitrary number of covariates. While under correct specification Type I error control will remain guaranteed, negative effects on the power should be expected. In this setting, we investigate the extent of such effects empirically, by including two additional continuous covariates on top of the proxy for drug exposure. However, while three covariates are included, the third one will in fact be irrelevant, i.e.~independent of the resulting responses.

\begin{table}[t]
    \centering
    \resizebox{0.7\textwidth}{!}{
    \setlength{\tabcolsep}{0.5em} 
    \renewcommand{\arraystretch}{1.35}
    \centering
    \begin{tabular}{c| c c |c c} \toprule
        Ground truth $\eta$ & $\tau$ & Size of $S_\tau$ & GLM (logistic, additive) & ISS \\ \toprule
        Logistic model (additive) & 0.5 & 0.96 & \checkmark & \checkmark\\ 
        Non-smooth \& non-monotone & 0.35 & 0.87 & & \\ \bottomrule 
    \end{tabular}}
    \caption{See the caption of Table~\ref{table:simulations_AE_univariate_table_of_functions} and the text for details.}
    \label{table:simulations_AE_multiv_table_of_functions}
\end{table}

\begin{figure}[t]
     \centering
     \begin{subfigure}[c]{0.24\textwidth}
        \centering
        \includegraphics[width=0.83\textwidth]{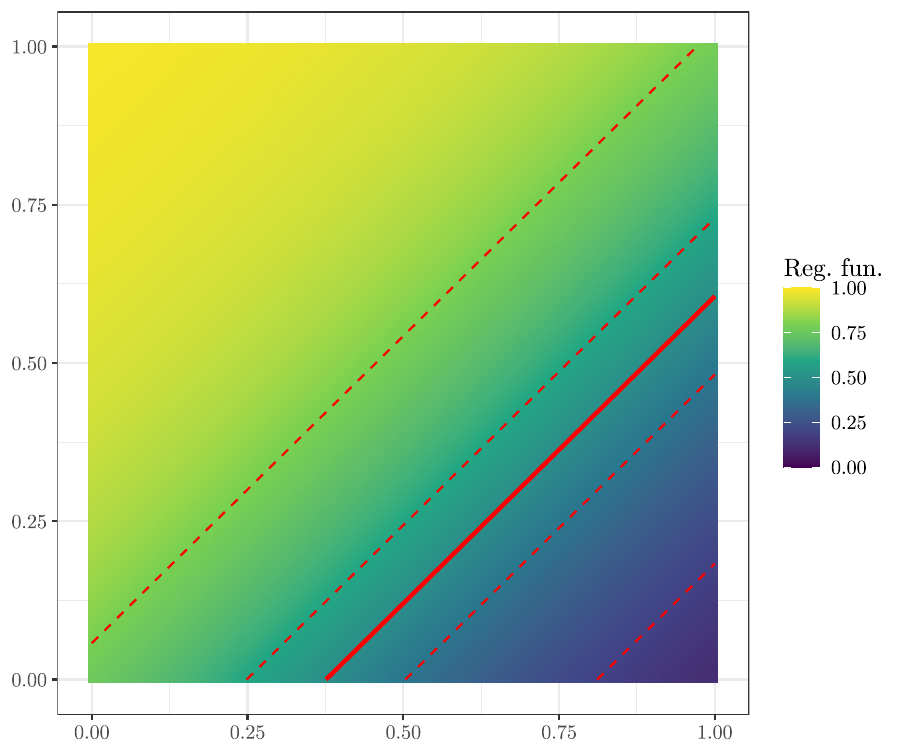}\vspace{0.32cm}
    \end{subfigure}
    \begin{subfigure}[c]{.75\textwidth}
        \centering
        \includegraphics[width=\textwidth]{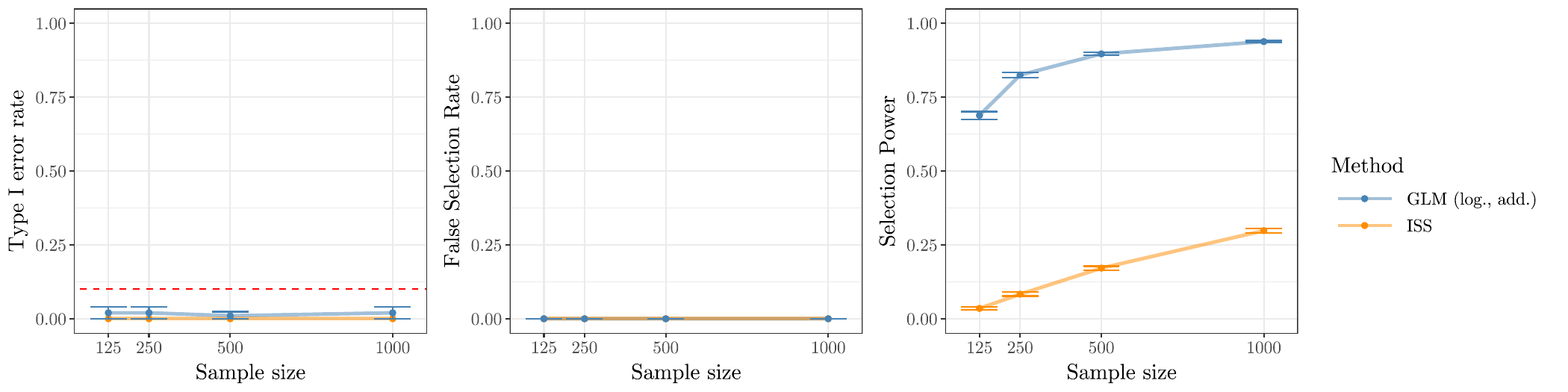}
    \end{subfigure}%
    \\ \vspace{0.2cm}
    \begin{subfigure}[c]{0.24\textwidth}
        \centering
        \includegraphics[width=0.83\textwidth]{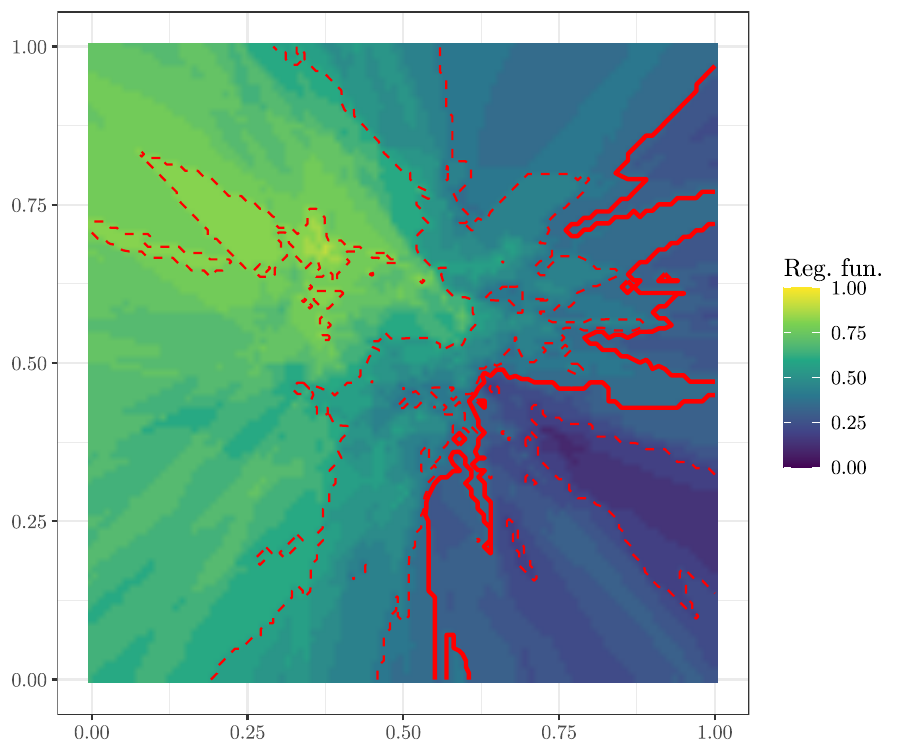}\vspace{0.32cm}
    \end{subfigure}
    \begin{subfigure}[c]{.75\textwidth}
        \centering
        \includegraphics[width=\textwidth]{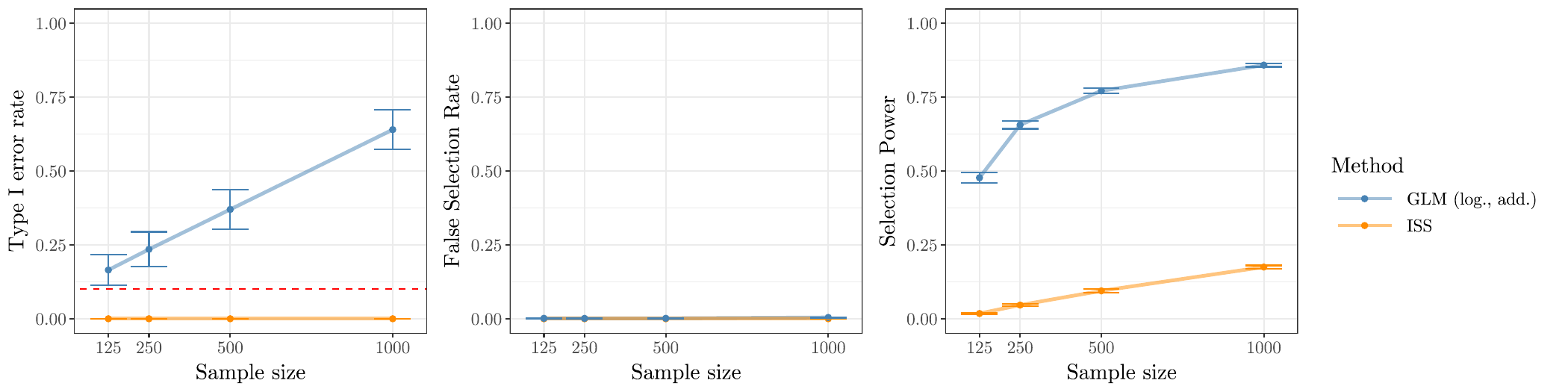}
    \end{subfigure}%
        \caption{The left-most column gives the ground truth regression function $\eta$ of the ground truths (``Logistic model (additive)'' in the top row and ``Non-smooth \& non-monotone'' in the bottom row), where the axes represent the first and second covariate (with the third not affecting the value of $\eta$). The colour represents the value of $\eta(x)$ at the corresponding point and the red dashed lines give the contour line at level $0.2$, $0.4$, $0.6$ and $0.8$ respectively. The solid red line gives the contour line at the level $\tau$ and hence the boundary of $S_\tau$. The other columns show the empirical performance. See the caption of Figure~\ref{fig:AE_logmod_results} for details.}
        \label{fig:AE_multiv_results}
\end{figure}

\paragraph{Setup}
We consider the data-generating ground truth models described in Table~\ref{table:simulations_AE_multiv_table_of_functions}, which are extensions of the correspondingly named settings from the previous subsection. While $d=3$ here, only the first two covariates are relevant, so that we may still visualise $\eta$, as is done in Figure~\ref{fig:AE_multiv_results}. We see that for ``Logistic model (additive)'', ISS is correctly specified by supposing a negative effect of the first and third covariate, but a positive effect of the second one, which is how the method will be applied. As in previous sections, the direction of the effects in the ground truth ``Non-smooth \& non-monotone'' are roughly going in the same direction, but no monotonicity (let alone linearity) is present, so that neither of the two considered methods is correctly specified.

\paragraph{Results}
In the scenario ``Logistic model (additive)'' (top row of Figure~\ref{fig:AE_multiv_results}), the GLM is correctly specified, so that the GLM-based method controls the Type I error rate, while achieving very high power. In the second scenario (bottom row of Figure~\ref{fig:AE_multiv_results}), where the GLM-based method is misspecified, we observe severe violation of the limit set on the Type I error rate, although the false selection rate remains extremely close to $0$. ISS on the other hand controls both the false selection rate and the Type I error rate, but selects smaller subgroups  (see top row of Figure~\ref{fig:AE_multiv_results}), illustrating the effect of the curse of dimensionality.

\subsection{Discovering the subgroup with high treatment effect}
\label{sec:simulations_HTE}
We now continue with a simulation study on the setting described in Section~\ref{sec:application_clinical_trials_TEH}. Here, the subgroup of interest contains all patients whose covariate configurations cause them to have a conditional average treatment effect (CATE) of at least $\threshold$, as described in Section~\ref{sec:methods_hte}. There are two key aspects different in this setting compared to that of Section~\ref{sec:simulations_AE}. Firstly, while the responses in the previous section were binary and hence, conditional on the concomitant covariate observations, Bernoulli-distributed, they are now continuous, which allows for much more complex distributions. Secondly, we are now working in a counterfactual setting in which each patient will be assigned to either treatment or control.

\subsubsection{Data generation}
We use the \textbf{benchtm} package in \texttt{R} \citep{sun2024benchtm} to generate simulated datasets. 
This package allows for the creation of data with varying degrees of treatment effect heterogeneity, 
designed to mimic real clinical trial data. The distributions of the simulated variables are based on those 
from a Phase~III study of a compound targeting an inflammatory disease (see \cite{sun2024benchtm} for more details).  

In our setup, we generate samples of size $n \in \{125, 250, 500, 1000\}$, each consisting of $d = 30$ biomarkers. 
Among these,~8 are categorical and 22 are continuous. All biomarkers are standardized to lie within the unit interval $[0,1]$. To generate continuous responses, the package \textbf{benchtm} offers four different scenarios, that differ in the form of the prognostic contribution of the covariates $f_\mathrm{prog}(x) := \E(Y_i\bigm| T_i = 0, X_i = x)$ and of the respective CATE. We summarise the different scenarios in Table~\ref{table:benchtm_scenarios}. Given any one of these four scenarios, we then generate the response $Y_i$ for each $i \in \{1,\ldots,n\}$ independently by first sampling the treatment $T_i \sim \mathrm{Bin}(0.5)$ independently of $X_i$, before letting $Y_i := f(X_i,T_i) + \varepsilon_i$ for independent $\varepsilon_i \sim \mathcal{N}(0,1)$, where 
\[
f(x, t) := f_\mathrm{prog}(x) + t\cdot \bigl(\beta_0 + \beta_1\cdot f_{\mathrm{pred}}(x)\bigr)
\]
for $t \in \{0,1\}$, $x\in \R^d$ and $\beta_0$, $\beta_1$, $f_\mathrm{prog}$ and $f_\mathrm{pred}$ specified in Table~\ref{table:benchtm_scenarios}. We repeat this sampling procedure independently $B = 100$ for reliability of our simulation results.

\begin{figure}[tp]
     \centering
     \begin{subfigure}[c]{.23\textwidth}
         \centering
        \includegraphics[width=\textwidth]{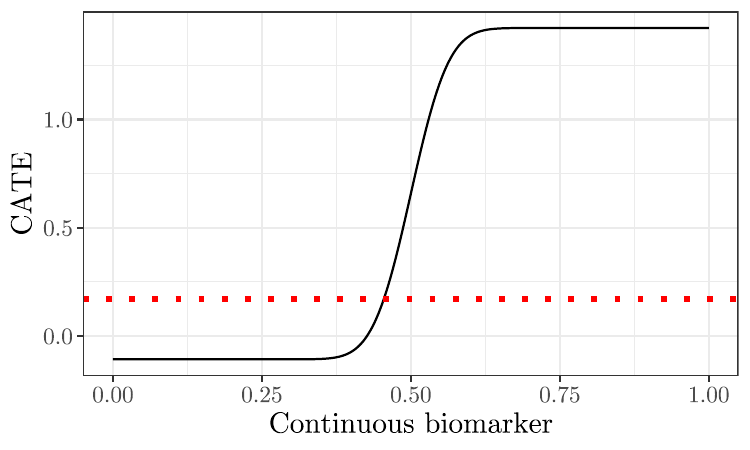}
     \end{subfigure}
    \begin{subfigure}[c]{.75\textwidth}
        \centering
        \includegraphics[width=\textwidth]{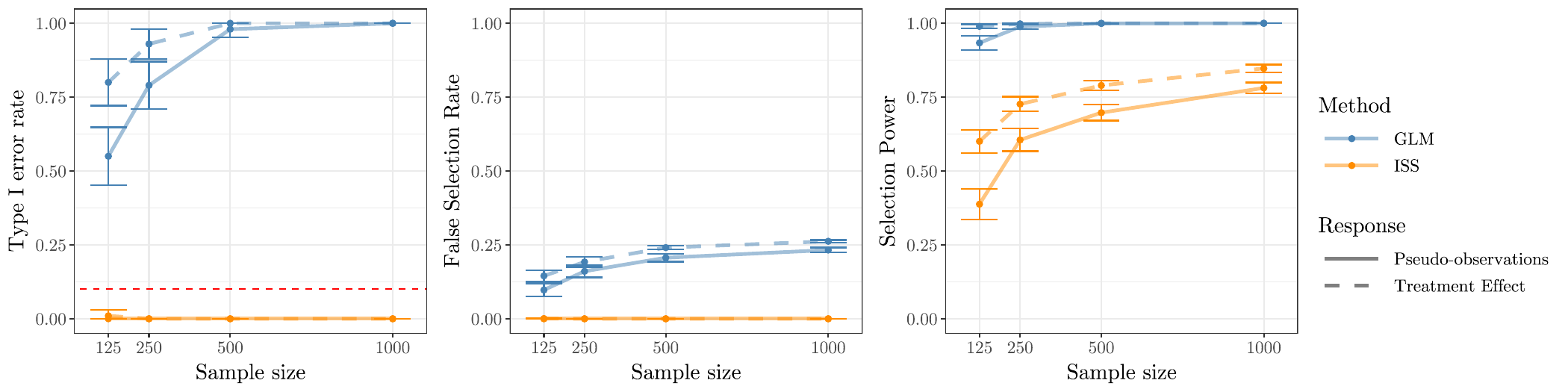}
    \end{subfigure}%
    \\ \vspace{0.2cm}
    \begin{subfigure}[c]{.23\textwidth}
         \centering
        \includegraphics[width=\textwidth]{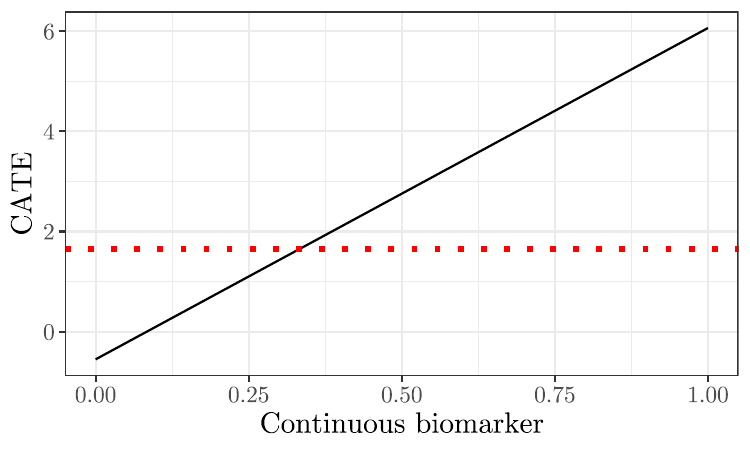}
     \end{subfigure}
    \begin{subfigure}[c]{.75\textwidth}
        \centering
        \includegraphics[width=\textwidth]{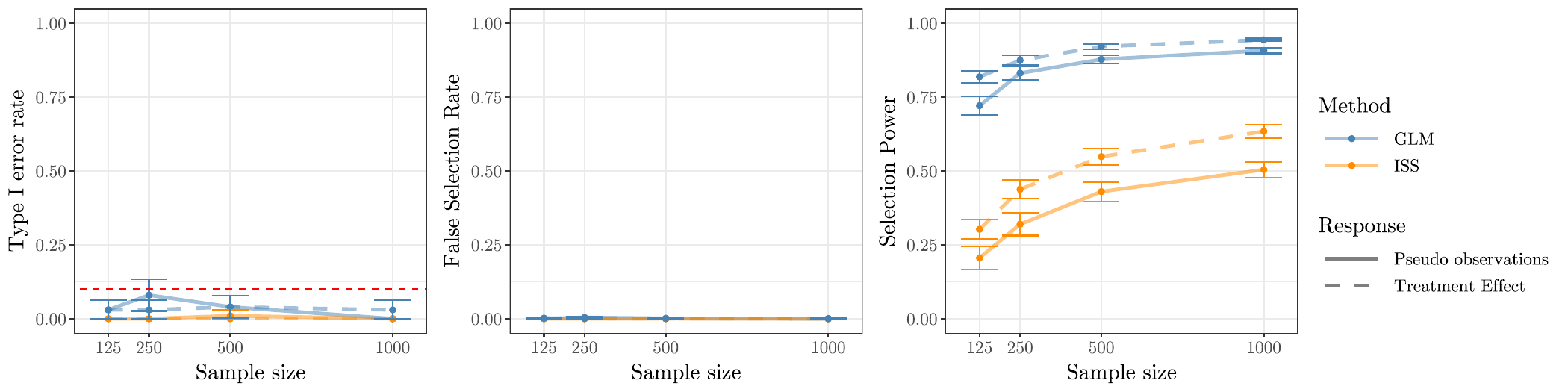}
    \end{subfigure}%
     \\ \vspace{0.2cm}
     \begin{subfigure}[c]{.23\textwidth}
         \centering
        \includegraphics[width=\textwidth]{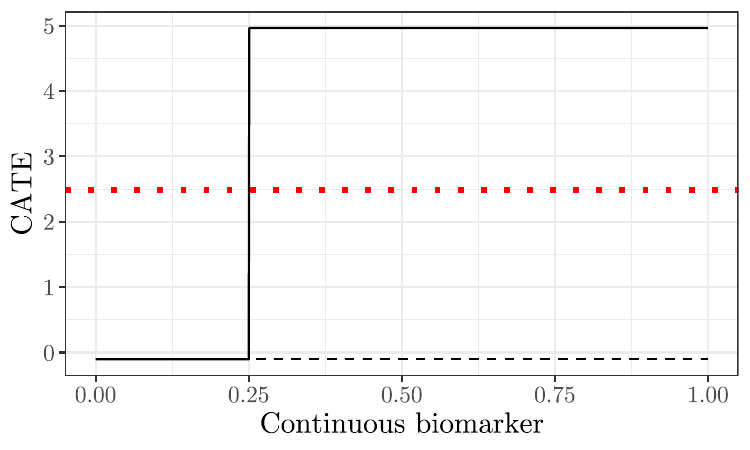}
     \end{subfigure}
    \begin{subfigure}[c]{.75\textwidth}
        \centering
        \includegraphics[width=\textwidth]{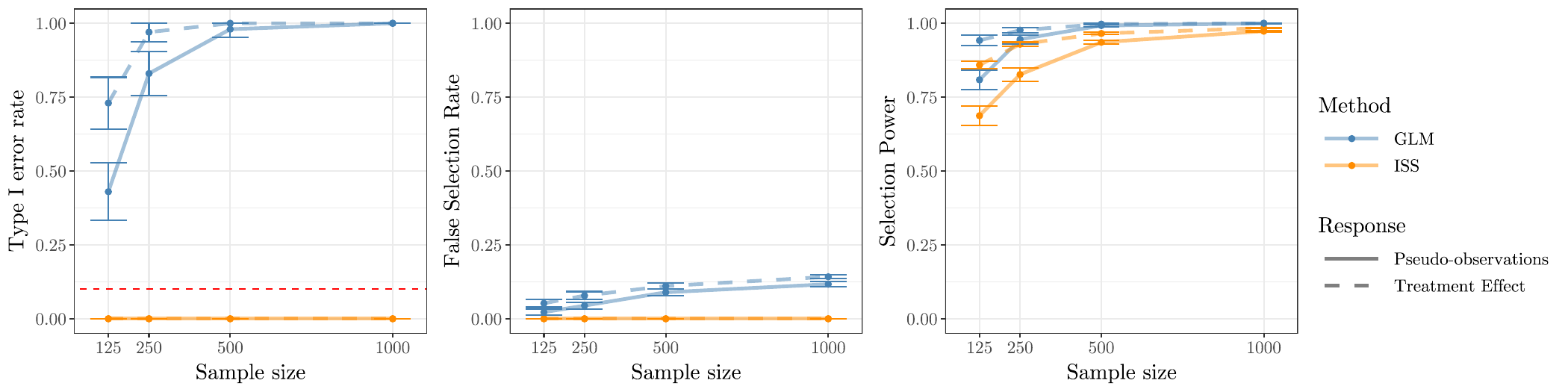}
    \end{subfigure}%
     \\ \vspace{0.2cm}
     \begin{subfigure}[c]{.23\textwidth}
         \centering
        \includegraphics[width=\textwidth]{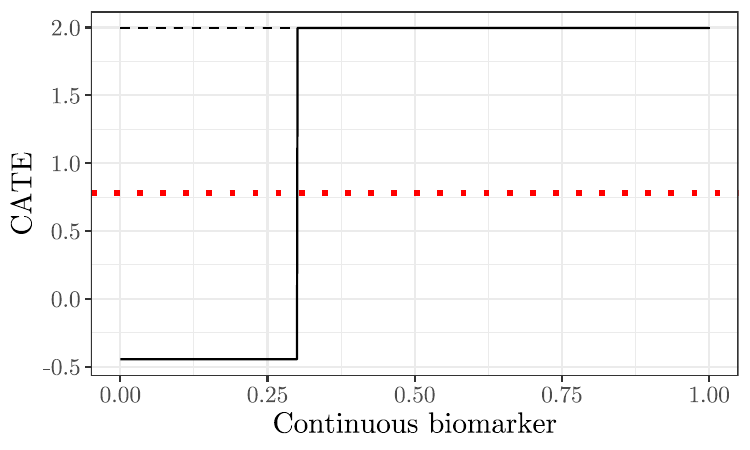}
     \end{subfigure}
    \begin{subfigure}[c]{.75\textwidth}
        \centering
        \includegraphics[width=\textwidth]{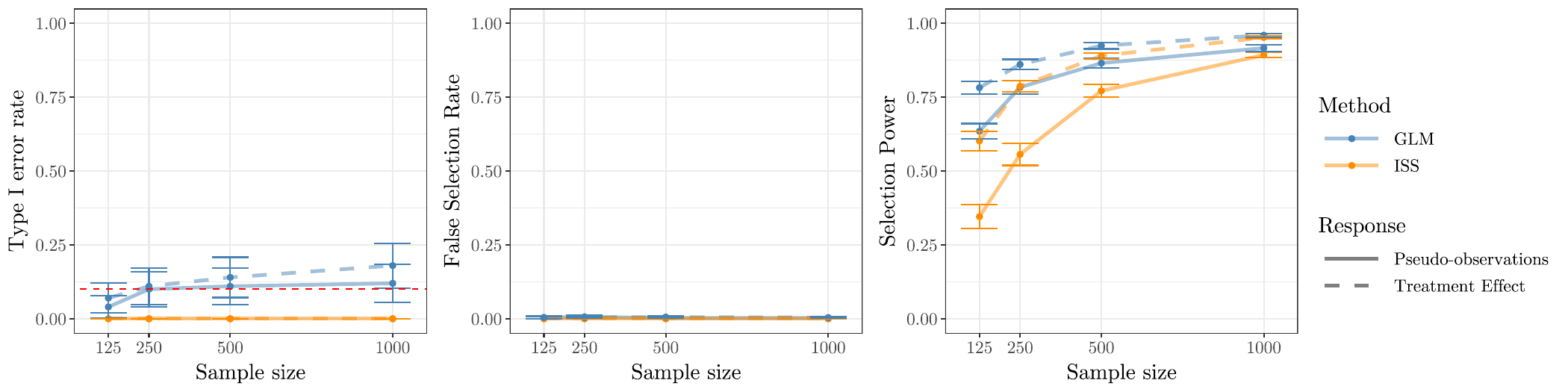}
    \end{subfigure}
        \caption{Each row of panels corresponds to one scenario defined in Table~\ref{table:benchtm_scenarios}; the first row gives results for the scenario labelled ``Gaussian CDF'', the second for ``Linear'', the third for ```And'-condition'' and the fourth for ```Or'-condition''. The first column gives the corresponding function $\mathrm{CATE}(x)$, where the $x$-axis represents the continuous variable on which it depends, and should it also depend on a categorical variable (bottom two rows), the effect of that is visualised through the linetype. The other three columns give simulation results for, in order, the Type I error rate, the false selection rate and the selection power of the considered methods.}
        \label{fig:HTE}
\end{figure}

\redCommentToBeRemoved{(This line break is new.)} Note that the CATE is given as the second summand of the above formula; i.e.~$\mathrm{CATE}(x) = \beta_0 + \beta_1 \cdot f_{\mathrm{pred}}(x)$ and hence we take the scenario labels to describe $f_\mathrm{pred}$. Inspection of the scenarios in Table~\ref{table:benchtm_scenarios} illustrates that the CATE will be driven by at most 2 variables, even if 30 covariates are recorded (and in part are associated with the response). We illustrate the different ground truths of the CATE in the left-most column of Figure~\ref{fig:HTE}.

\begin{table}[t]
    \resizebox{\textwidth}{!}{
    \setlength{\tabcolsep}{0.5em} 
    \renewcommand{\arraystretch}{1.35}
    \centering
    \begin{tabular}{c|c|c|cc|cc} \toprule
                Scenario & $f_{\mathrm{pred}}(x)$ &  $f_\mathrm{prog}(x)$ & $\beta_0$ & $\beta_1$ & $\tau$ & Size of $S_\tau$ \\ \hline
                Gaussian CDF  & $\Phi\bigl(20\bigl\{x^{(11)} - 0.5\bigr\}\bigr)$ & $2.3 \cdot \bigl(0.5\mathbbm{1}\bigl\{x^{(1)} = 1\bigr\} + x^{(11)}\bigr)$ & $-0.11$ & $1.53$ & $0.17$ & $0.52$ \\
                Linear  & $x^{(14)}$ & $1.41\cdot \bigl(x^{(14)} - \mathbbm{1}\bigl\{x^{(8)} = 0\bigr\}\bigr)$ & $-0.55$ & $6.62$ & $1.65$ & $0.51$\\
                `And'-condition  & $\mathbbm{1}\bigl\{x^{(14)} > 0.25 \text{ and } x^{(1)} = 0\bigr\}$ & $1.38 \cdot \bigl(\mathbbm{1}\bigl\{x^{(1)} = 0\bigr\} - 0.5x^{(17)}\bigr)$ & $-0.10$ & $5.07$ & $2.48$ & $0.46$\\
                `Or'-condition & $\mathbbm{1}\{x^{(14)} > 0.3 \text{ or } x^{(4)} = 1\}$ & $2.9\cdot \bigl(x^{(11)} - x^{(14)}\bigr)$ & $-0.45$ & $2.44$ & $0.78$ & $0.81$\\ \bottomrule
    \end{tabular}}
    \caption{We write $x = \bigl(x^{(1)}, \ldots, x^{(30)}\bigr)^\top$, where only the elements $x^{(1)}, x^{(4)}, x^{(8)}, x^{(11)}, x^{(14)}$ and $x^{(17)}$ play a role. Out of these, $x^{(1)}, x^{(4)}, x^{(8)}$ take values in $\{0,1\}$ while all others are continuous on $[0,1]$. The values of $\beta_0$ and $\beta_1$ are rounded to two digits.}
    \label{table:benchtm_scenarios}
\end{table}

\paragraph{Individual treatment effects}
Beyond the above data generation process, we also consider --- as a point of reference --- the application directly to individual treatment effects sampled independently for each patient as $\mathrm{CATE}(X_i) + \epsilon_i$ for $\epsilon_i \sim \mathcal{N}(0,2)$, with $X_i$ and $\epsilon_i$ independent. Note that this quantity is what we would observe if we had access to the difference of the responses under $T_i = 1$ and $T_i = 0$, each with their own independent standard Gaussian noise. Since we can only assign a patient to either control ($T_i = 0$) or treatment ($T_i = 1$) in practice, this is a very idealized scenario, that we only include to investigate how much the fundamental problem of only observing one of the outcomes influences the performance of each method.

\subsubsection{Estimand: the target subgroup}
To choose a suitable value of $\tau \in \R$ to define the subgroup $S_\tau = \{x \in \R^d: \mathrm{CATE}(x) \geq \tau\}$, we followed a similar approach as described in Section~\ref{sec:simulations_AE_target}. Here, $\mathrm{CATE}(x)$ takes the role of the regression function $\eta(x)$ and we use the covariate distribution given by \textbf{benchtm} as the distribution under which we ensure $S_\tau$ to be of reasonable size. The choice of $\tau$ as well as the resulting measure of $S_\tau$ are given in Table~\ref{table:benchtm_scenarios}. 

\subsubsection{Methods}\label{sec:simulations_HTE_methods}
We now give details on the high-level description of the doubly robust learner method described in Section~\ref{sec:methods_hte}. Here, we let $X_i$ denote the $i$th covariate vector, where we only include relevant variables, i.e.~all those that are either prognostic or predictive.

We generate the pseudo-outcome for a particular $i\in \{1,\ldots,n\}$ by first splitting our data into $K \in \mathbb{N}$ folds, where we let $k(i) \in \{1,\ldots, K\}$ denote the fold of the data containing observation $i$. Then, based on the data in all folds except the $k(i)$th, we train estimators for $(\pi, \Tilde{\regressionFunction}^0, \Tilde{\regressionFunction}^1)$, which we denote by $(\hat{\pi}_{-k(i)}, \hat{\regressionFunction}^0_{-k(i)}, \hat{\regressionFunction}^1_{-k(i)})$ using a super-learner framework combining GLMs with elastic net penalization and conditional random forests using the \texttt{R}-package \textbf{SuperLearner} \citep{polley2024SuperLearner}. We then output the pseudo-observation
\[
    \Tilde{Y}_i := \hat{\regressionFunction}^1_{-k(i)}(X_i) - \hat{\regressionFunction}^0_{-k(i)}(X_i) + \frac{T_i - \hat{\pi}_{-k(i)}(X_i)}{\hat{\pi}_{-k(i)}(X_i)\bigl(1 - \hat{\pi}_{-k(i)}(X_i)\bigr)}\bigl(Y_i - \hat{\regressionFunction}^{T_i}_{-k(i)}(X_i)\bigr).
\]

Repeating this for all $i \in \{1,\ldots,n\}$ (where we only need to construct the estimators $K$ times), we generate pseudo-observations $\tilde{Y}_1, \ldots, \tilde{Y}_n$. For the purpose of our simulations, we have taken $K = 4$.

To the data-sets consisting of these pseudo-observations together with the predictive covariates for the respective setting, we then apply the methods discussed in Section~\ref{sec:methods}. More specifically, we apply the GLM-based method described in Section~\ref{sec:methods_glm} under the assumption of a Gaussian linear regression model. For ISS, we specify the direction of the relationships correctly and assume symmetric errors, making the $p$-value construction described in Section~\ref{sec:methods_ISS_pvalue_quantile} applicable. Note that the assumption of symmetry of $\tilde{Y}_i$ about its $X_i$-conditional expectation may be unfounded --- we are not claiming in this section that our methods are correctly specified, but rather seek to characterize their behavior for the difficult task of selecting a large portion of the patient subgroup with high CATE.

\subsubsection{Performance metric}
We evaluate our methods' ability to reliably and efficiently detect the subgroup of patients with high enough CATE via the metrics presented in Section~\ref{sec:simulations_AE_metrics}.

\subsubsection{Simulation results}
The simulation results are presented in Figure~\ref{fig:HTE}. These results are in line with the findings in Section~\ref{sec:simulations_AE}, as they illustrate that the GLM-based method can violate Type~I error rate control severely when the model is misspecified.  ISS does indeed control Type~I error in line with its proven guarantee, though it may select smaller subgroups than the GLM-based method under correct specification of the model for the latter method.  Interestingly, it seems that the power of the GLM-based method suffers less than ISS when applying the procedures to the pseudo-observations rather than the (in practice unobservable) individual treatment effects.

\section{Revisiting the application in clinical trials}\label{sec:application_clinical_trials_revisiting}
We now return to the two case studies presented in Section \ref{sec:application_clinical_trials}. Using the methods discussed above, we aim to identify the subgroups of interest within each context.

\subsection{Revisiting case study 1 (safety): Subgroup selection for the absence of adverse reactions}
\label{sec:application_clinical_trials_AE_revisiting}

In this section, we revisit the first case study to evaluate the performance of the proposed subgroup selection methods. Due to confidentiality purposes, we do not disclose the actual covariates used in the analysis.  The dataset comprises 247 patients, of whom 27 experienced the adverse event of interest. To demonstrate the potential of the proposed methods, we focus on two covariates of mixed type: a continuous variable $ X^{(1)} $, known to be relevant to the occurrence of adverse events, i.e. the higher the value the more likely to develop an AE, and a binary variable $ X^{(2)} $, representing a demographic characteristic that is known to be irrelevant. The inclusion of $ X^{(2)}$ serves to demonstrate the methods' ability to handle mixed-type covariates. Subgroups will be defined based on these two variables. We scale the continuous variable $ X^{(1)} $ to take values within the interval $[0, 1]$.

Revisiting our research question, we aim to identify subpopulations for which the probability of not experiencing the adverse event, conditional on covariates $ X^{(1)}, X^{(2)} $, exceeds a predefined safety threshold $ \tau $. This condition can be formally expressed as: $ \Pr(Y = 1 \mid X^{(1)}, X^{(2)}) \geq \tau,$
while ensuring control of the Type I error rate. For both methods under comparison, we set the safety threshold for the probability of not experiencing an AE to $ \tau = 0.8 $, and the nominal Type I error rate to $ \alpha = 0.05 $. The objective is to identify patient subgroups for which the conditional probability of remaining AE-free exceeds the threshold, thereby indicating that post-dose monitoring may not be necessary.

\begin{figure}[tbp]
    \centering
    \begin{subfigure}{.49\textwidth}
    \includegraphics[width=\linewidth]{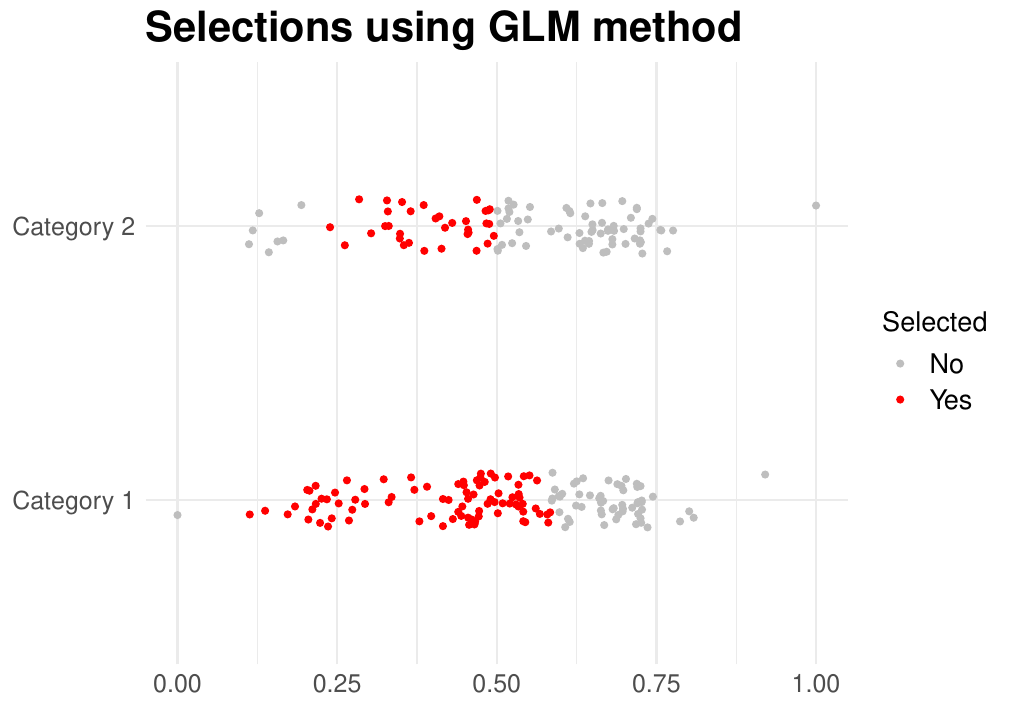}
    \end{subfigure}%
    \begin{subfigure}{.49\textwidth}
    \includegraphics[width=\linewidth]{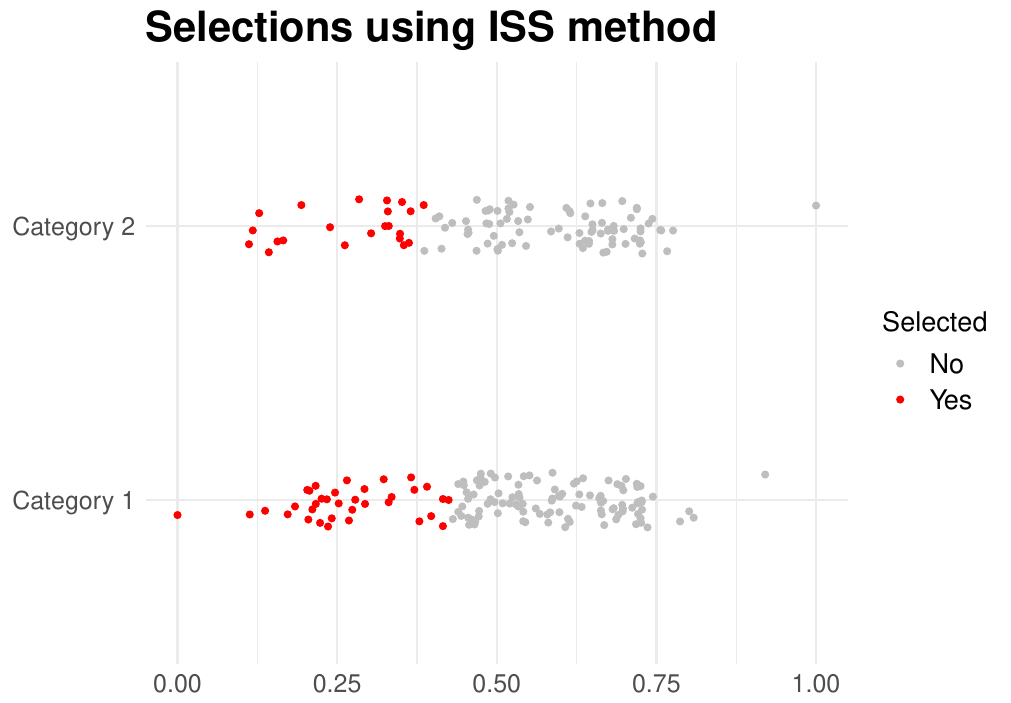}
    \end{subfigure}%
    \caption{Subgroup selection results for Case Study 1 (safety) using the GLM-based method (left) and the ISS method (right), when the cut-off value is $ \tau = 0.8$ and the nominal Type I error rate $ \alpha = 0.05.$ Each dot represents a patient, and a red dot indicates that the patient was selected by the method. The $x$-axes give the values of $X^{(1)}$, while the $y$-axes correspond to the binary variable $X^{(2)}$.}
    \label{fig:case_study_1}
\end{figure}

From the results shown in Figure~\ref{fig:case_study_1}, we observe that both methods consistently select patients with low values of $X^{(1)}$, while $X^{(2)}$ appears to have limited relevance, as the selections are similar across its categories. This aligns with our expectations, given that high values of $X^{(1)}$, which can be interpreted as a proxy for exposure, are known to increase the likelihood of adverse events. In contrast, $X^{(2)}$ represents a demographic variable that has not shown any influence on the occurrence of the adverse event.

Interestingly, the GLM-based method leads to a larger number of selected subgroups compared to the ISS method.  This implies that ISS certifies safety for fewer patients, suggesting that more patients should remain under post-dose monitoring. While this approach may appear more conservative, our simulation study demonstrated that it offers greater robustness in controlling the Type I error rate. As a result, it reduces the likelihood of false positives --- i.e., incorrectly identifying a patient as not requiring post-dose monitoring when they may in fact experience an adverse event, serving as a critical safeguard in this application.

\subsection{Revisiting case study 2 (efficacy): Subgroup selection for desired treatment response} 
\label{sec:application_clinical_trials_TEH_revisiting}

\begin{figure}[ht]
    \centering
    \begin{subfigure}{.49\textwidth}
    \includegraphics[width=\linewidth]{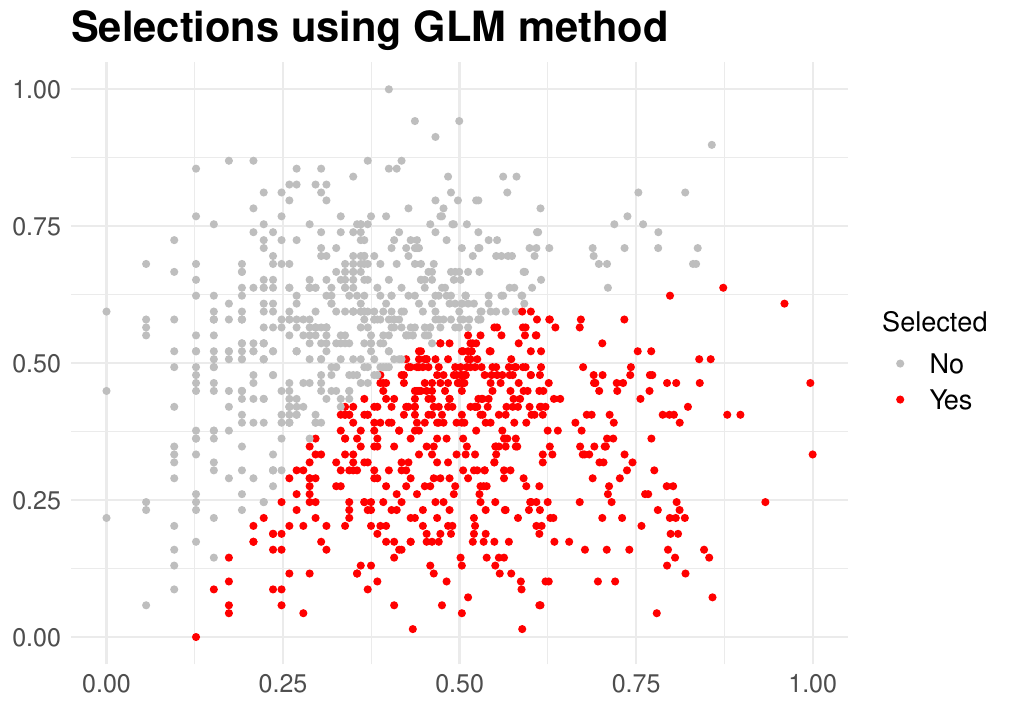}
    \end{subfigure}%
    \begin{subfigure}{.49\textwidth}
    \includegraphics[width=\linewidth]{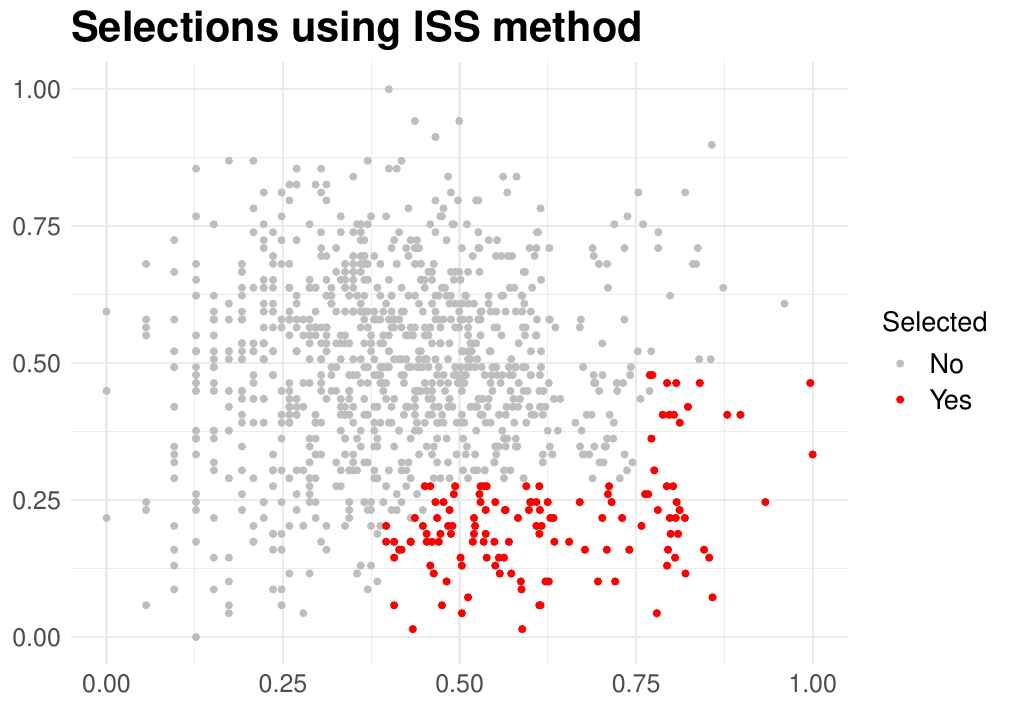}
    \end{subfigure}%
    \caption{Subgroup selection results for Case Study 2 (efficacy) using the GLM-based method (left) and the ISS method (right), when the cut-off value is $ \tau = 10$ and the nominal Type I error rate $ \alpha = 0.10.$ Each dot represents a patient, and a red dot indicates that the patient was selected by the method. The $x$-axes give the values of $X^{(1)}$, while the $y$-axes correspond to the binary variable $X^{(2)}$.}
    \label{fig:case_study_2}
\end{figure}
The second case study aimed to identify subgroups of psoriatic arthritis patients who experience a high treatment effect in response to Cosentyx, and we refer to these patients as exceptional responders. We focus on one of the approved doses of Cosentyx, 300 mg. Our dataset includes 1067 patients, with 444 receiving Cosentyx and 623 assigned to placebo. We define exceptional responders as individuals for whom the expected difference in the endpoint value at week 16 between Cosentyx ($T=1$) and placebo ($T=0$) exceeds a predefined threshold of $\tau$. To show the generality of the considered subgroup selection methods, in this section we will focus on a continuous endpoint, in contrast with the previous section where the endpoint (adverse event) was binary. This endpoint ($Y$) is a self-reported score of overall disease activity, recorded on a scale from 0 to 100.

Subgroups were defined using two continuous variables that had been identified as promising effect modifiers in previous studies \citep{Sechidis2021,bornkamp2023predicting}. As noted earlier, the goal of these case studies is to demonstrate the utility of the considered methods, rather than to make claims about treatment efficacy. Importantly, because our analysis is post-hoc and the variables included were identified as promising effect modifiers from previous works that used the same data, it is not reasonable to expect any formal control of the Type I error rate. We refrain from using the original names of these variables and will refer to them as $X^{(1)}$ and $X^{(2)}$, scaling their values to lie within the interval $[0, 1]$. Furthermore, we know the monotonicity of the effects; the larger the $X^{(1)}$ the higher the effect, while the opposite holds for $X^{(2)}$, the lower the higher the effect. 

The estimand of interest in our example is the mean difference, which can be expressed as: $E(Y \mid T = 1, X^{(1)},X^{(2)}) - E(Y \mid T = 0, X^{(1)},X^{(2)}) \ge \tau, $ with threshold $ \tau = 10 $ that is selected for illustrative purposes to represent a high treatment effect on the mean difference scale. This implies that we want to identify the subgroup of patients where the expected response on the patients that take the experimental drug is at least 10 units higher than those receiving placebo. To ensure statistical rigor, we controlled the {Type I error rate} at a significance level of $\alpha = 0.10,$ a level widely used for subgroup analysis \citep{dane2019subgroup}.

Figure~\ref{fig:case_study_2} presents the subgroups identified using the two methodologies discussed in this paper. First of all, we can see that both of the variables are important in identifying the subgroups, since the methods select patients with high values of $X^{(1)}$ and low values of $X^{(2)}$. By the strong reliability guarantees for subgroup selection we demand in our framework, patients falling into the selected areas could be deemed as exceptional responders if the analysis here were not conducted post-hoc and if assumptions associated to the methods are satisfied. Furthermore, as earlier in the safety case study, the GLM-based method leads to a higher number of discoveries compared to ISS. Since we are considering a real dataset here and not a simulation, it is impossible to assess, whether this is due to higher power of the GLM-based method under correct specification or the more reliable Type I error rate control of ISS.

\section{Discussion and Conclusions}
\label{sec:conclusions} 
\subsection{Conclusions}
In this work, we present and evaluate data-driven methods for controlled subgroup selection. Although broadly applicable, we focus on clinical trials to illustrate their practical relevance --- particularly in two critical areas of drug development: (1) identifying patients at low risk of adverse events (safety), and (2) selecting subgroups of patients who exhibit high conditional treatment effects (efficacy). Both objectives are pursued with explicit control of a strong Type I error rate to ensure statistical rigour and reliability, far exceeding that of traditional approaches to subgroup analysis.

We investigate two complementary approaches: a GLM-based method that leverages parametric modelling and achieves higher power under correct model specification but leads to severe violation of the Type I error rate under model misspecification, and the ISS method, which is valid under the mild assumption of a monotonicity constraint. Extensive simulation studies alongside two case studies demonstrate the strengths and limitations of each method.

\subsection{Discussion on practical considerations}
Our work aligns with ongoing regulatory discussions. For instance, the recent FDA draft guidance \citep{FDA2025} on the use of artificial intelligence in regulatory decision making explicitly highlights subgroup identification in safety monitoring as a key motivating use case. By offering methods that combine flexibility with strong statistical guarantees, our framework addresses a critical regulatory need: enabling data-driven discovery while ensuring reliability and reproducibility. In safety focused applications, these guarantees help regulators build confidence that identified subgroups genuinely meet predefined safety thresholds, thereby reducing unnecessary monitoring without compromising patient protection. In efficacy focused applications, such guarantees can support claims of differential benefit across populations, an increasingly important consideration in labeling decisions and precision medicine initiatives.

While the methods discussed provide strong Type I error control under their stated assumptions, no approach is entirely immune to model misspecification. Assessing the sensitivity of conclusions across a broad spectrum of plausible scenarios is a critical component of rigorous analysis. Accordingly, we extensively evaluated diverse scenario configurations, and future work can expand this to cover an even broader range of plausible scenarios.

\sloppy Finally, for controlled subgroup selection methods to have meaningful impact in drug development, they must be embedded within structured, pre-planned workflows. The WATCH framework (Workflow to Assess Treatment Effect Heterogeneity) exemplifies such an approach, providing a systematic process that begins with prespecified analysis planning \citep{sechidis2025watchworkflowassesstreatment}. Once initial data analysis and preprocessing are complete, the methods presented in this paper can be applied to perform controlled subgroup selection in a principled and reproducible manner.

\subsection{Discussion on methodological considerations and future directions}

One natural direction for future research is to develop theoretical results on false selection rate control. Our current framework provides very strong guarantees at two separate layers; for instance when requiring that $\hat{L}_\alpha \subseteq S_\tau$ holds with probability at least $1 - \alpha$ when establishing treatment efficacy, we demand that \emph{every} point in $\hat{L}_\alpha$ has an expected treatment effect above $\tau$ and that this is the case \emph{with high probability}. There are also other weaker versions of such a guarantee that one could consider. For instance, weakening the former of these two aspects, we could require that $\hat{L}_\alpha$ is such that $\P\bigl(\P(X_0 \in \hat{L}_\alpha\setminus S_\tau \mid X_0 \in \hat{L}_\alpha, \hat{L}_\alpha) \geq 1 - q \bigr) \geq 1 - \alpha$  for some pre-specified tolerable false selection rate $q \in (0,1)$, where the inner probability is over the randomness in $X_0$ and the outer one over the randomness in the data used to construct $\hat{L}_\alpha$. In particular, whenever this is the case, i.e., $\P(X_0 \in \hat{L}_\alpha\setminus S_\tau \mid X_0 \in \hat{L}_\alpha, \hat{L}_\alpha) \geq 1 - q$, a proportion of $1-q$ of patients falling into $\hat{L}_\alpha$ have a conditionally expected response of at least~$\tau$, so we yield a guarantee on the $q$-quantile of the conditionally expected responses within the selected subgroup. While not as strong as guaranteeing a conditional expected treatment effect above the threshold for every covariate value in the selected subgroup, this may still be a more useful guarantee than the average expected treatment effect that is currently often considered when evaluating subgroups.

Our framework exploits specific modelling assumptions (e.g., linearity or monotonicity) to achieve non-trivial power while still controlling the Type I error rate for finite sample sizes. 
It would be of interest to establish (asymptotic) validity of methods building on the meta-learning approach discussed in this paper under suitable conditions. 

Finally, from a practical perspective, it is important to provide guidance on sample sizes at which controlled subgroup selection is realistic. Our simulations highlight the trade-off between sample size, model assumptions, and the achievable precision of subgroup identification. In this context, one may argue that for truly confirmatory conclusions, it is not sufficient to control only Type I error, but that control of Type II errors is also needed. Typically, the latter is being addressed at the trial design stage through appropriate sample size planning. This highlights the need for clear criteria for power and minimal detectable subgroup effects, so that valid subgroup selection may be embedded into the trial design framework in practice.

\section*{Acknowledgments}
The authors would like to thank Alain Marti, Jelle J.~Goeman and Rajen D.~Shah for their feedback on the manuscript.

\section*{Funding}
The research of MMM and RJS was supported by RJS's European Research Council Advanced Grant 101019498.

\bibliographystyle{apalike}
\bibliography{bibl.bib}

\end{document}